\documentclass[a4paper,11pt]{article}
\usepackage{jinstpub} 
\usepackage{enumitem}
\usepackage{verbatim}
\usepackage{lineno}

\title{A new method for structural diagnostics with muon tomography and deep learning}

\author[a]{Lorenzo Pezzotti}
\author[b]{Davide Cifarelli}
\author[c,d]{Daniele Corradetti}
\author[d]{José Paulo Costa}
\author[b]{Giorgio Gabrielli}
\author[e]{Lorenzo Galante}
\author[e]{Antonio Gallerati}
\author[f,g,h,i]{Ivan Gnesi}
\author[b]{Andrea Jouve}
\author[b,g]{Alessio Marrani}

\affiliation[a]{Istituto Nazionale di Fisica Nucleare, Sez. Bologna, viale Carlo Berti Pichat 6/2, 40127 Bologna, Italy}

\affiliation[b]{Elementar, Divisione Ricerca e Sviluppo, Galleria Enzo Tortora 21, 10121 Torino, Italy}

\affiliation[c]{Grupo de Física Matemática, Instituto Superior Técnico, Av. Rovisco Pais, 1049-001, Lisboa, Portugal}

\affiliation[d]{STAP Reabilitação Estrutural, SA Rua General Ferreira Martins 8 - 9B, Algés, 1495-137, Portugal}

\affiliation[e]{Politecnico di Torino, corso Duca degli Abruzzi 24, 10129 Torino, Italy}

\affiliation[f]{Centro Ricerche Enrico Fermi, piazza del Viminale 1, 00184 Roma, Italy}

\affiliation[g]{Dipartimento di Management Valter Cantino, Corso Unione Sovietica 218 Bis, 10134, Torino, Italy}

\affiliation[h]{INFN - Laboratori Nazionali di Frascati, Via Enrico Fermi 54, 00044 Frascati, Italy}

\affiliation[i]{CERN, 1211 Geneva 23, Switzerland}

\emailAdd{lorenzo.pezzotti@cern.ch}

\abstract{This work investigates the production of high-resolution images of typical support elements in concrete structures by means of muon tomography (muography). By exploiting detailed Monte Carlo radiation-matter simulations, we demonstrate the feasibility of reconstructing 1 cm-thick iron bars inside 30 cm-deep concrete blocks, regarded as an important testbed within the structural diagnostics community. In addition, we present a new method for integrating simulated data with advanced deep learning techniques in order to improve the muon imaging of concrete structures. Through deep learning enhancement techniques, this results in a dramatic improvement in image quality and a significant reduction in data acquisition time, which are two critical limitations within the usual practice of muography for civil engineering diagnostics.}

\keywords{Computerized Tomography (CT) and Computed Radiography (CR), Muon spectrometers, Simulation methods and programs, Image processing}

\makeatletter
\gdef\@fpheader{%
    JINST \textbf{20} (2025) P06034;\, \href{https://doi.org/10.1088/1748-0221/20/06/P06034}{10.1088/1748-0221/20/06/P06034}
     }
\makeatother

\begin{document}
\maketitle
\flushbottom

\section{Introduction}\label{sec:intro}
Muon imaging is a well-known non-invasive technique exploiting cosmic ray muons to investigate the internal structure of dense objects. Muons are elementary particles created when cosmic rays interact with atoms in the Earth's atmosphere, generating showers of secondary particles which, traveling at nearly the speed of light, reach the Earth's ground and further cross wide thicknesses of matter. This continuous shower of muonic particles makes them highly effective for probing dense materials, usually impenetrable to other types of radiation, such as X-rays \cite{zhang2020muography}.

The main idea at the core of muon imaging is to detect the attenuation or scattering of muons as they  move through a medium. As intuitively reasonable, denser regions, i.e. regions with a higher density of the budget material,  give rise to a greater attenuation or scattering, which can then be mapped in order to reveal the internal structure of the traversed objects.  Remarkably, this crucial feature of muonic fluxes have  found application in a variety of contexts, ranging from archaeology \cite{alvarez1970search,morishima2017discovery,cimmino20193d},  to geophysics \cite{tanaka2009detecting,tanaka2013subsurface,tioukov2019first}, nuclear waste management \cite{borozdin2003radiographic,thomay2016passive}, and civil engineering
\cite{niederleithinger2022muon}.%
\footnote{%
For a complete review  and a list of references, see e.g.\ \cite{IAEA2012}.}

Within the field of civil engineering, the diagnostics of concrete structures provides a natural framework for the application of muon tomography (usually named muography).  In fact, this context is plagued by the increasing issue of the aging of concrete structures: as time passes, bridges, tunnels, and historic buildings constructed throughout the 20th century are all reaching the end of their planned lifetime \cite{wang2021time, chen2021life}.
A faithful assessment of the state of concrete structures is often hindered by the lack of complete or accurate technical documentation on the structure themselves; the original construction projects are often missing, so that even basic information on the original structural configuration and matter composition is not available
\cite{structural2000guideline,aci2019assessment}. Unfortunately,
traditional probing methods, such as ground-penetrating radar (GPR), ultrasound techniques and infrared thermography, are mostly ineffective and inaccurate when dealing with dense and thick materials, like reinforced concrete \cite{maierhofer2003nondestructive,verma2013review}. Indeed, GPR typically penetrates the concrete until a depth of at most 2 meters only, and it has major issues when dealing with multiple reinforcement layers. On the other hand, ultrasonic methods  work fine for detecting cracks, but quickly become ineffective as the thickness of the concrete increases. Finally, infrared thermography can provide information about the state of surface layers, but says nothing about possible internal voids or defects.

The aforementioned problems can be effectively dealt with by muography, especially for what concerns the depth of penetration and energy efficiency \cite{tanaka2023muography}. However, of course, muography comes with its own limitations: while being more penetrative, it can suffer from poor spatial resolution, caused by the scattering of muons. For instance, the horizontal resolution
of typical muon imaging systems is strongly constrained by the geometry and angular acceptance of the detectors \cite{tanaka2009detecting}, whereas the vertical resolution suffers from quite severe limitations due to the natural angular distribution of cosmic ray muons, which suffer from a horizontal bias causing anisotropic imaging artifacts \cite{niederleithinger2022muon}.  Also, an important limit of muography is the time required to reach a number of detections of muons large enough to grant for images with acceptably high resolution. In fact, the muon flux at sea level is approximately 170 particles per square meter per second \cite{particle2020review}, while, within standard techniques, millions of muons are needed to generate statistically meaningful images \cite{niederleithinger2022muon}. This results in data acquisition times that range from days to weeks and makes muography too cumbersome and essentially unpractical; for example, diagnosing an infrastructure such as a bridge would require a service interruption of several days, which is usually not an option.

In order to address the problems described above, in the present paper we propose a new setup integrating
\textsc{Geant4}, the state-of-the-art simulation toolkit for particle interactions with matter \cite{agostinelli2003geant4,allison2006geant4,allison2016recent}, along with an image enhancement through deep learning which exploits standard data segmentation techniques combined with U-Nets with residual dense blocks.  On the one hand, by means of muonic simulations, we will investigate the spatial resolution issue within the analysis of reinforced concrete structures. On the other hand, by training neural networks on simulated datasets, we will be able to significantly reduce the amount of muon events needed to reconstruct a meaningful image.\smallskip

The paper is developed according to the following structure. Section \ref{sec:methodology} introduces the methodology, presenting the integration of \textsc{Geant4} simulations with neural network models. Section \ref{sec:Geant4}  deals with the design and implementation of the simulations, providing a detailed treatment of the key parameters, as well as of the analytical results obtained.  Then, Section \ref{sec:neural}  pertains to the development of neural network algorithms, focusing on architecture, data augmentation techniques, and training procedures. A detailed discussion of the results, including both quantitative and qualitative evaluations of the reconstructed images, is provided in Section \ref{sec:results}. Finally, in Section \ref{sec:conclusion} we conclude with a summary of our findings, with an outlook of potential directions for future researches and applications.

\section{Methodology overview}
\label{sec:methodology}

The methodology proposed in this paper makes use of \textsc{Geant4} simulations for modeling the interaction of muons with reinforced concrete, generating synthetic data to train neural networks aimed at enhancing image resolution and reducing noise.

As the first step, detailed \textsc{Geant4} simulations are developed and executed, in order to model the passage of muons through concrete structures containing iron reinforcements. For this study, our preliminary objective was to determine the feasibility of identifying iron cylinders with a diameter of 10 mm embedded in a $30\times30\times30$ cm$^3$ concrete cube, arranged in various configurations. This geometric setup was considered fundamental by our engineering consultants, in view of further specialized developments in structural diagnostics. Clearly, different practical scenarios -- such as bridge diagnostics with beams and pre-stressing cables, water tanks, and underwater structures -- will require different simulations to generate synthetic data aligned with the structures under examination. For the present preliminary feasibility study, muons are generated in the simulation with energies between 3 and 4 GeV, uniformly distributed over the top surface of the concrete block. Detection planes,  placed up to 1 m upstream and downstream of the block, recorded the positions of the muons before and after interacting with the sample. Then, the deflection angles of the muon tracks undergo a detailed analysis, in order to infer the presence and position of the iron cylinders within the concrete.

Once a large simulation dataset pertaining to the structure being diagnosed has been produced, the second step is to develop the deep learning models to improve the data obtained from the simulations. In this investigation, our priority is to reconstruct high-resolution images from datasets with a limited number of muon events. This can be achieved by applying U-Net architectures enriched with residual-in-residual dense blocks (RRDB), known for their effectiveness in super-resolution tasks \cite{wei2023esrgan,chen2024training}. Although reducing the number of muon events was our primary goal, as demonstrated in Section \ref{sec:results}, the proposed architecture also achieves a significant and visible reduction in the signal-to-noise ratio.

Once the network architecture is established, the third step consists of the use of the simulated data from \textsc{Geant4} as input for the neural network training. These datasets are subjected to data augmentation techniques, including rotations, reflections and flips, in order to expand the training data and to improve the model generalization.

Thus, by integrating analytical reconstruction techniques with deep learning models, our methodology achieves two key results  while bringing muography into civil engineering practice. Firstly, it allows for the reconstruction of internal iron structures even from datasets with a limited number of muon events: as shown in Section \ref{sec:results}, inference on images derived from $10^5$ muon events yields results comparable to images derived from $4.5\times10^6$ events. Secondly, the background noise resulting from the event variability gets naturally smoothed: this feature is explicitly measured by the Gradient Magnitude Similarity Deviation (GMSD) metric when compared to Ground Truth, as discussed in Section \ref{sec:results}.\par\smallskip

All in all, this work demonstrates that the combination of \textsc{Geant4} simulations and deep learning techniques gives rise to a remarkably effective approach to the enhancement of the spatial resolution and accuracy of muography, paving the way to a number of potential applications in non-invasive monitoring of complex engineering structures.

\section{\textsc{Geant4} simulations and analytical results}\label{sec:Geant4}

\subsection{Simulation design and strategy}
To address the possibility of reconstructing the metallic support structure inside concrete elements, we performed detailed and comprehensive simulations of the interaction of negatively charged muons in several sample structures. To this aim, we made use of the \textsc{Geant4} software (release 11.1.0); this is a multi-purpose simulation toolkit for Monte Carlo simulations of radiation-matter interactions. In particular, \textsc{Geant4} allows for the tracking of any particle in any material with great accuracy, taking into account the vast majority of interaction processes from the TeV scale down to thermal energies. For these reasons, \textsc{Geant4} is currently a standard for simulations of particles interacting with matter in nuclear, high-energy, biomedical and atmospheric physics. The complete simulation software to study the passage of muons across reinforced concrete structures was designed as a C\texttt{++} code on top of the \textsc{Geant4} library.

In this project, our main goal was to investigate the possibility of detecting and resolving metal structures immersed in concrete blocks. The concrete block had a common geometry, namely a $30\times30\times30$ cm$^3$ cube. The composition of the concrete was a mixture of the following elements and mass fractions: H ($1.0\%$), C ($0.1\%$), O ($52.91\%$), Na ($1.6\%$), Mg ($0.2\%$), Al ($3.39\%$), Si ($33.70\%$), K ($1.39\%$), Ca ($4.40\%$), Fe ($1.4\%$). Moreover, the concrete density was set at $2.3$ $\mathrm{g/cm^3}$ and its radiation length ($X_0$) to $11.55$ cm. The concrete block was placed at the origin \,$\mathcal{O}=(0,0,0)$\, of a standard right-handed reference frame, and it was considered to be immersed in the Earth's atmosphere, at standard temperature, pressure and density.

\begin{figure}[!ht]
\centering
\fbox{\includegraphics[width=0.45\textwidth]{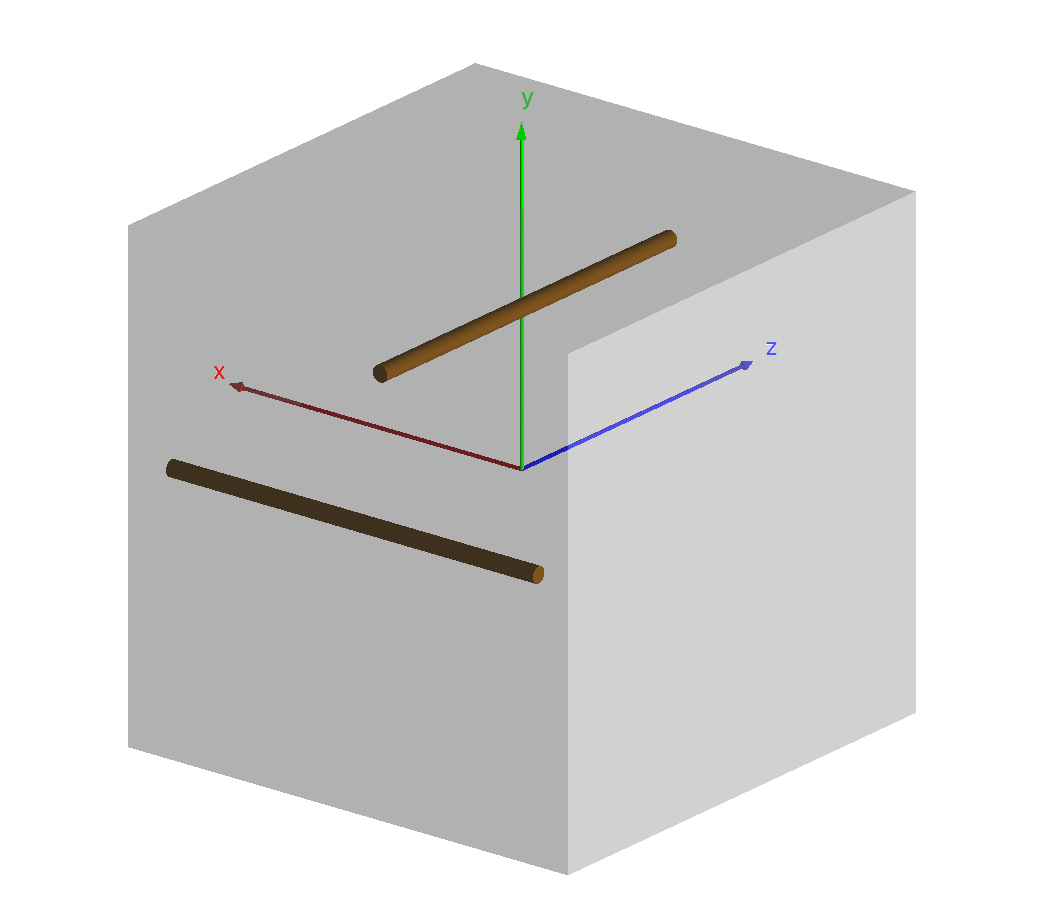}
\includegraphics[width=0.45\textwidth]{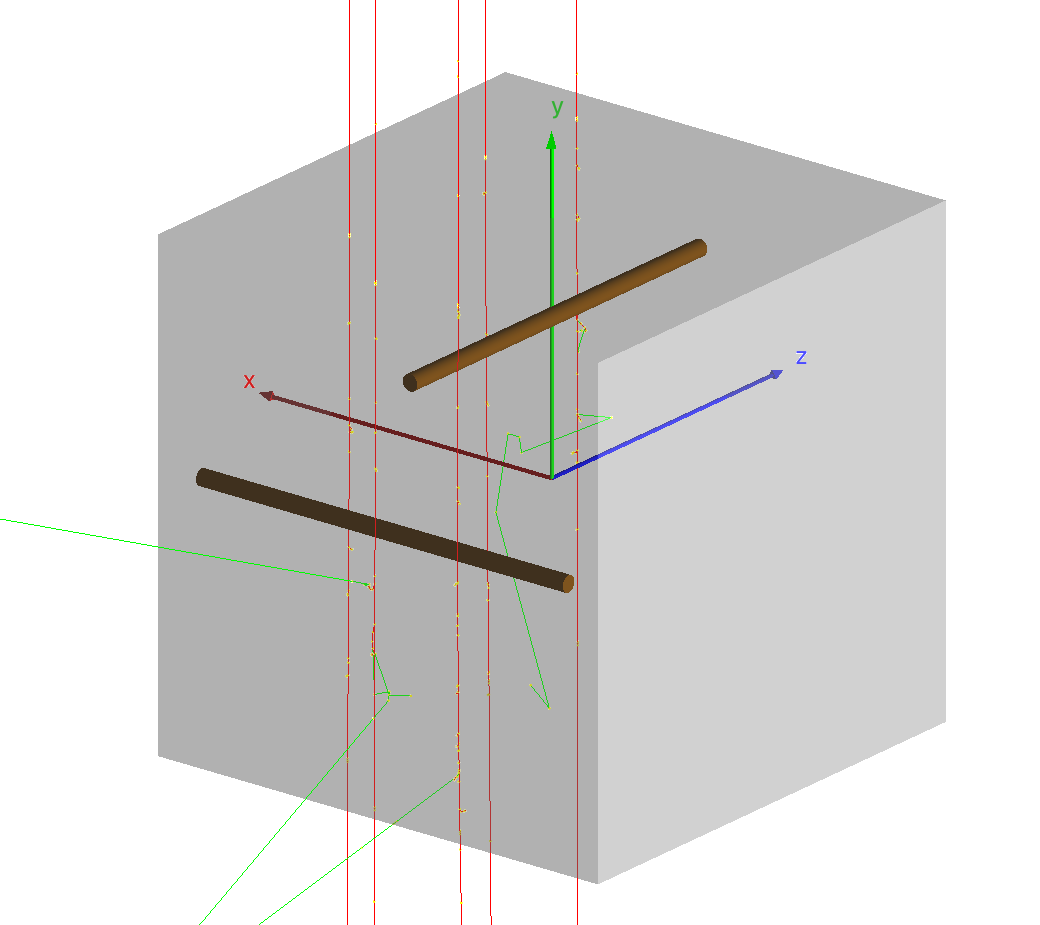}}
\caption{A simulated sample geometry, consisting of a concrete {block} housing two iron bars (left). Simulated interactions of five muons (red tracks), including the generation of secondary particles (green tracks) in the same geometry sample. Image from the \textsc{Geant4} simulation.}
\label{Geo1_1}
\end{figure}

Inside the concrete block, cylindrical iron bars of 1 cm diameter and 25 cm length were placed in random positions, parallel to the $x$ or $z$ axes, while the $y$ axis was the vertical one along which muons travel. In this framework, the density of the iron was set to $7.874$ g/cm$^3$, and its radiation length to $1.757$ cm. By placing a random number of bars from 0 to 15 inside the concrete block, a sample geometry was defined; for instance, Fig.~\ref{Geo1_1} (left) shows a sample geometry, in which two iron bars are inserted. Then, we simulated a total of 100 sample geometries and, for each geometry, we verified the absence of non-physical overlaps that would result in impossible intersections among the bars. This provided us with a set of geometries, which we were subsequently going to employ for analytical and machine learning studies.

For what concerns muons, we simulated a muon radiation source consisting of negatively charged muons, traveling along the direction \,$\texttt{Dir}=(0,-1,0)$\, and uniformly distributed over the face of the concrete block parallel to the $xz$ plane, with positive $y$ coordinate. To prevent muons from entering the block structure too close to its edges and thus possibly undergoing a scattering out of the sample, the impact points of muons were distributed on a surface of $28\times28$ cm$^2$ at the geometric center of the front face of the concrete block. Furthermore, the kinetic energy of the muons was simulated to be uniformly distributed in the range $3-4$ GeV, similar to the case in which the sample is exposed to accelerator particle beams. Fig.~\ref{Geo1_1} (right) provides a graphical representation of the interaction of five muon tracks (in red) within one sample geometry.

Our simulations tracked each primary muon through the sample, thus recording one event per primary muon. The tracking of muons in the simulation took into account every relevant physical process involved into muonic interactions within matter, like ionization, multiple scattering, bremsstrahlung, and lepto-nuclear interactions, as they are modelled within the FTFP\vbox{\hrule width.5em}\,BERT \textsc{Geant4} Physics List, which is recommended for high-energy physics simulations.

In order to record the muon position before and after its interaction within the sample, two detecting planes, parallel to the $xz$ plane, were included upstream and downstream the concrete element for a total of four detection points. Each detector area was $30\times30$ cm$^2$, identical to the concrete block area. The first layer was positioned 1 m before the samples's front face and the second was positioned 1 mm close to the sample's front face. Similarly, two layers were positioned downstream at 1 mm and 1 m distance to the sample's back face. When a muon track propagated through such planes, the $(x,z)$ coordinates were stored in files for later analysis. In the case in which the muon track did not reach the downstream detector plane, for instance because it induced a nuclear breakup in the sample geometry via the lepto-nuclear effect, we decided to neglect the event within our data analysis.

\begin{figure}[ht!]
\centering
\fbox{\includegraphics[width=0.5\textwidth]{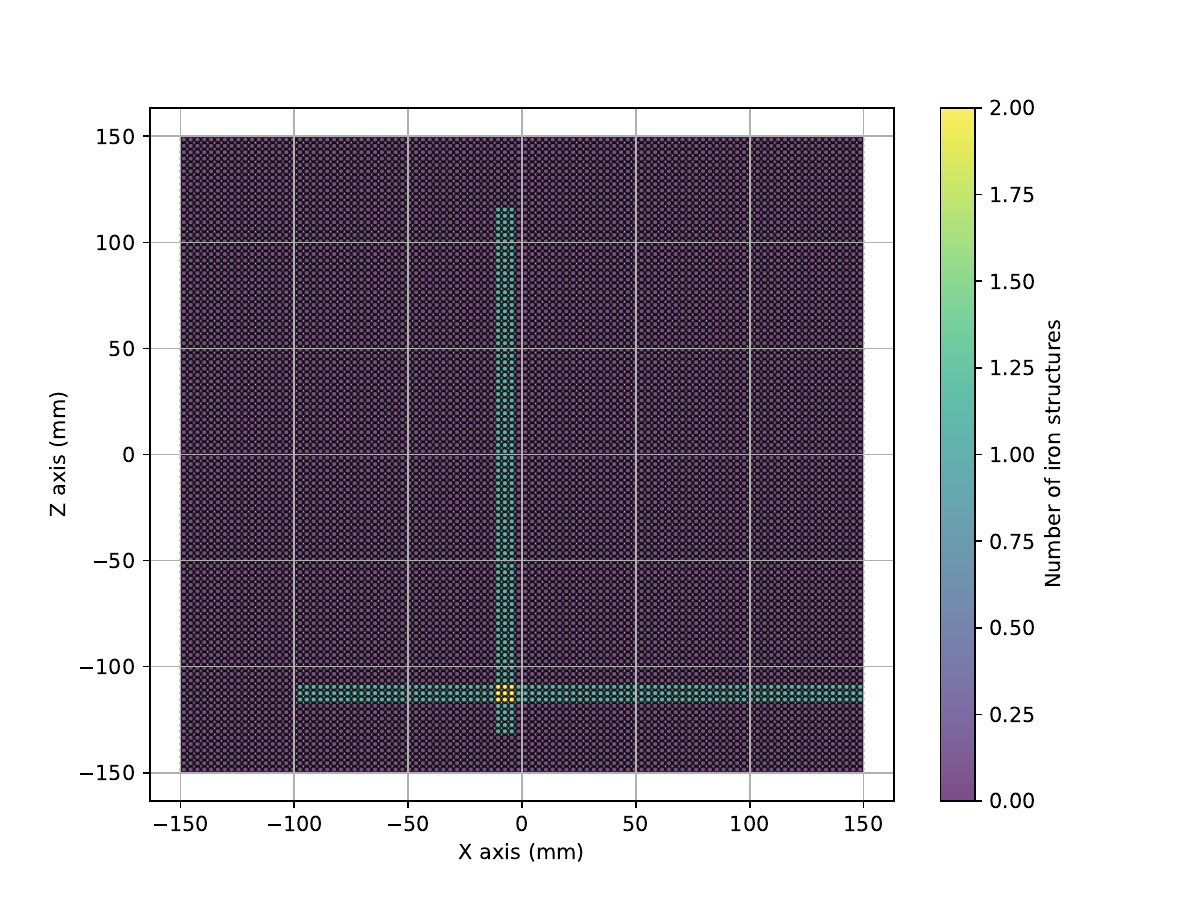}
\includegraphics[width=0.5\textwidth]{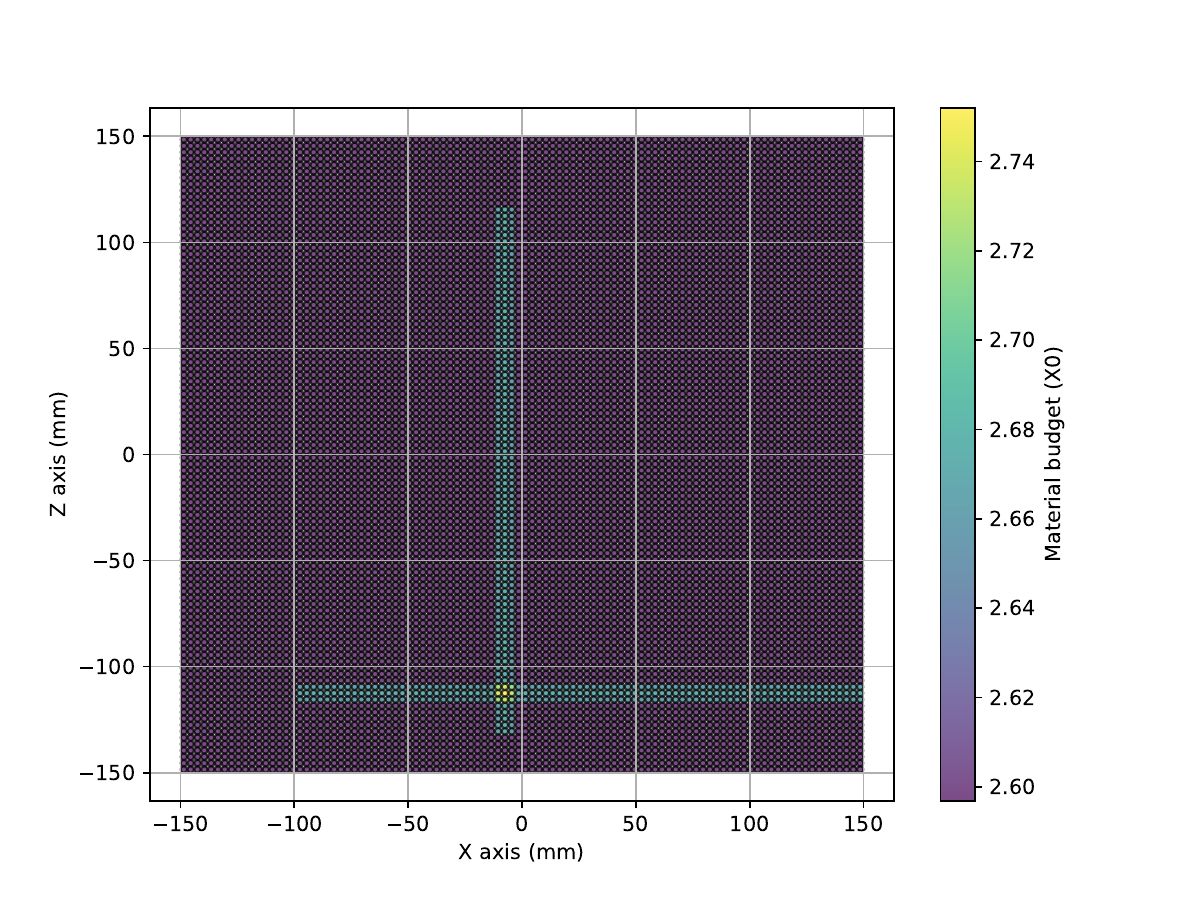}
}
\caption{Projections along {the} $y$-direction of the iron bars content inside the sample geometry shown in Fig.~\ref{Geo1_1} (left), {and of} the effective budget material in radiation lengths $X_0$ (right).}
\label{Geo1_1_Projections}
\end{figure}

The dataset employed for our investigation consisted of $18\times10^6$ events for each sample geometry, thus delivering a realistic set of synthetic data to address the possibility of resolving metallic support structures within concrete structures. In the case in which the muons traveled along the $y$-direction, only the projective image along this direction could be resolved, resulting in a two-dimensional figure in the $xz$ plane. Fig.~\ref{Geo1_1_Projections} shows the actual $y$-projection for the sample shown in Fig.~\ref{Geo1_1}, and it represents the target for the subsequent analyses. Indeed, it shows the number of bars inside the concrete structure (left) and the corresponding material budget map in radiation lengths $X_0$ (right). The relative budget material increases from 2.6 $X_0$ for the areas covered by concrete only, to 2.68 $X_0$ ($+3\%$) for the areas covering only one iron bar, to 2.75 $X_0$ ($+6\%$) for the areas covering two iron bars.

\subsection{Data analysis and results}
The Monte Carlo data stored for offline analysis consisted of vectors, containing the position of the primary muon track crossing detection layers. For the following analytical results, only the 1 m far upstream detector plane vector ($\texttt{Pos}_1$), and the 1~m far downstream detector plane vector ($\texttt{Pos}_2$) were used. The properties of the secondary particles produced by the muonic interaction within the samples were not considered in the analysis.

Several variables describing the deflection angle on muons passing through the sample under test can be used to reconstruct the budget material traversed. In our case, the first quantity exploited for the resolution of the metallic structures was the root mean square (RMS) of the muon tracks deflection angle. Within the energy range under consideration, this angle is almost uniquely due to the multiple scattering of charged particles within the electric field of the nuclei in the concrete block. In order to determine it, after calculating the vector $\texttt{Track}=\texttt{Pos}_2-\texttt{Pos}_1$, we computed its angle with the original muon directions $\texttt{Dir}=(0,-1,0)$ as the arc-cosine of the dot product of the normalized $\texttt{Track}$ and $\texttt{Dir}$ vectors. The resulting angle is therefore the scattering angle experienced by muons over a 2.3 meter distance: 1 meter of air plus 0.3 m of concrete, with an additional meter of air downstream.

\begin{figure}[!ht]
\centering
\fbox{\includegraphics[width=1.0\textwidth]{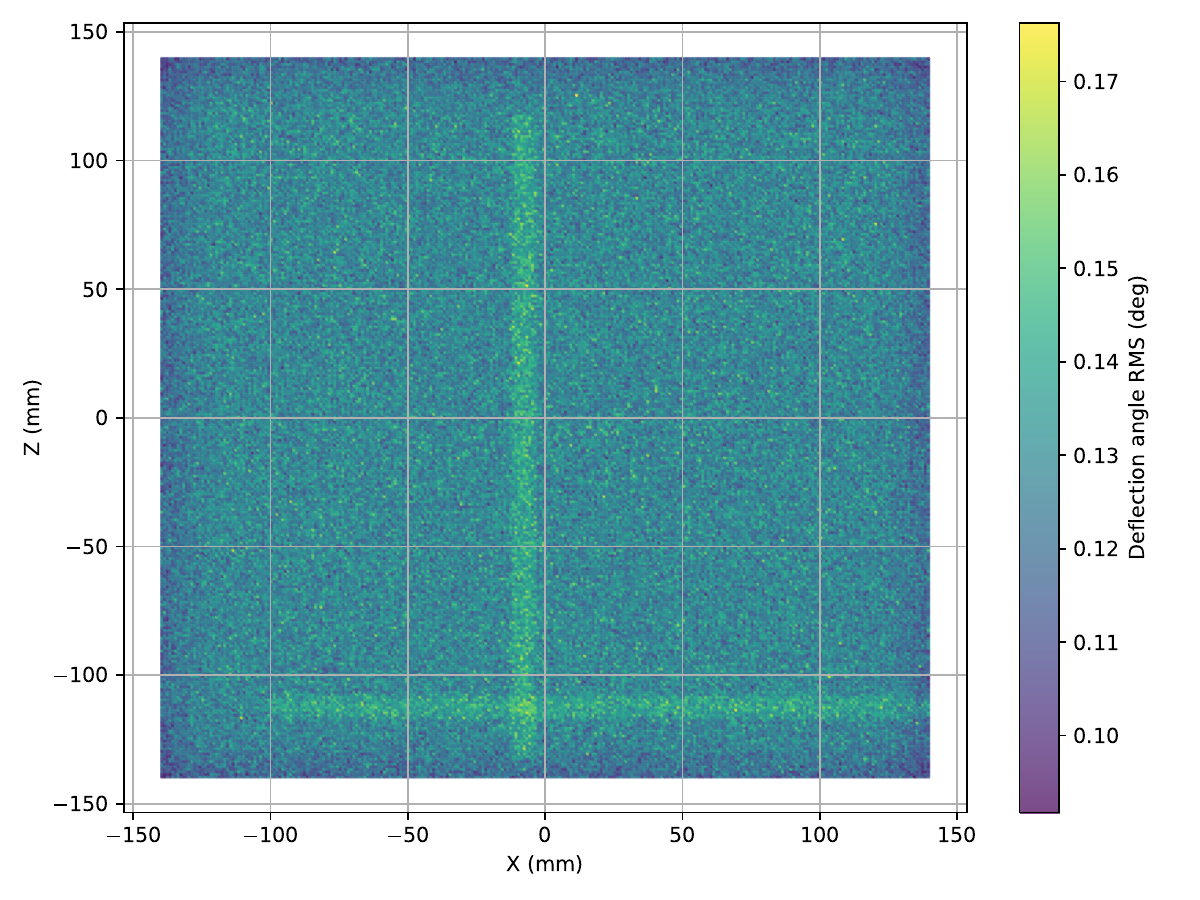}}
\caption{Scatter plot of the deflecting angle RMS of muon tracks passing through the sample geometry shown in Fig.~\ref{Geo1_1} and travelling along the $y$-direction. The image has been created by grouping muon tracks into $1$ mm$^2$ pixels.}
\label{Geo1_1_res0p0}
\end{figure}

We divided the front face of the sample geometry into $1\times1$ mm$^2$ pixels and collected the distributions of the scattering angle for muons entering each pixel separately. This procedure defined $280\times280$ pixels, corresponding to an equal number of angular distributions. While using $18\times10^6$ muon events per sample, each pixel was crossed on average by $230$ muons. Per each pixel, we extracted the scattering angle RMS from the corresponding distribution. The result for the geometry sample of Fig.~\ref{Geo1_1} is shown in Fig.~\ref{Geo1_1_res0p0}, and it should be compared with the true $y$-projection of the sample shown in Fig.~\ref{Geo1_1_Projections}: it is easy to appreciate the effectiveness of muon tomography for high-resolution images of metallic structures within concrete blocks, and its potentially successful applicability in monitoring civil engineering constructions.

Two important limitations of the approach outlined above will be considered in the following sections. We will discuss the impact of the detector spatial resolution on the imaging capability in Subsection~\ref{subsec:detector}, whereas in Section~\ref{sec:neural} we will discuss the need to expose the sample under study to a very high number of muon events in order to achieve high-resolution images.

Note that to create the images needed to train the neural network that improves our muographic results, we decided to compute the scattering angle experienced inside the concrete block only. To do that, the scattering angle was computed between the direction vector upstream and downstream the sample. They were obtained from the difference in position between the two upstream and downstream planes, respectively. Hence, the need to have four detector planes in our simulation.

\subsection{Relevance of the detector spatial resolution}
\label{subsec:detector}
The treatment outlined above did not take into account a realistic resolution in the detection of muon tracks passing through the upstream and downstream detecting layers. In order to address the importance of the spatial resolution of such detectors, we repeated the data analysis while applying a uniform smearing on the \texttt{Pos}$_1$ and \texttt{Pos}$_2$, $x$ and $z$ coordinates. This smearing was applied by drawing a uniformly distributed random number within $\pm$ the expected detector resolution and subsequently adding it to the measured coordinate. Four independent random numbers were generated for each event, according to a hypothetical resolution, and applied to the data. We considered resolutions of $5$, $10$ and $20$ mm. It should be noted that the impact of the detector resolution on the imaging results is twofold, since it introduces an error in the association of the muon track to the correct pixel and it also worsens the resolution on the measurement of the muon scattering angle.

\begin{figure}[!ht]
\centering
\fbox{\includegraphics[width=0.5\textwidth]{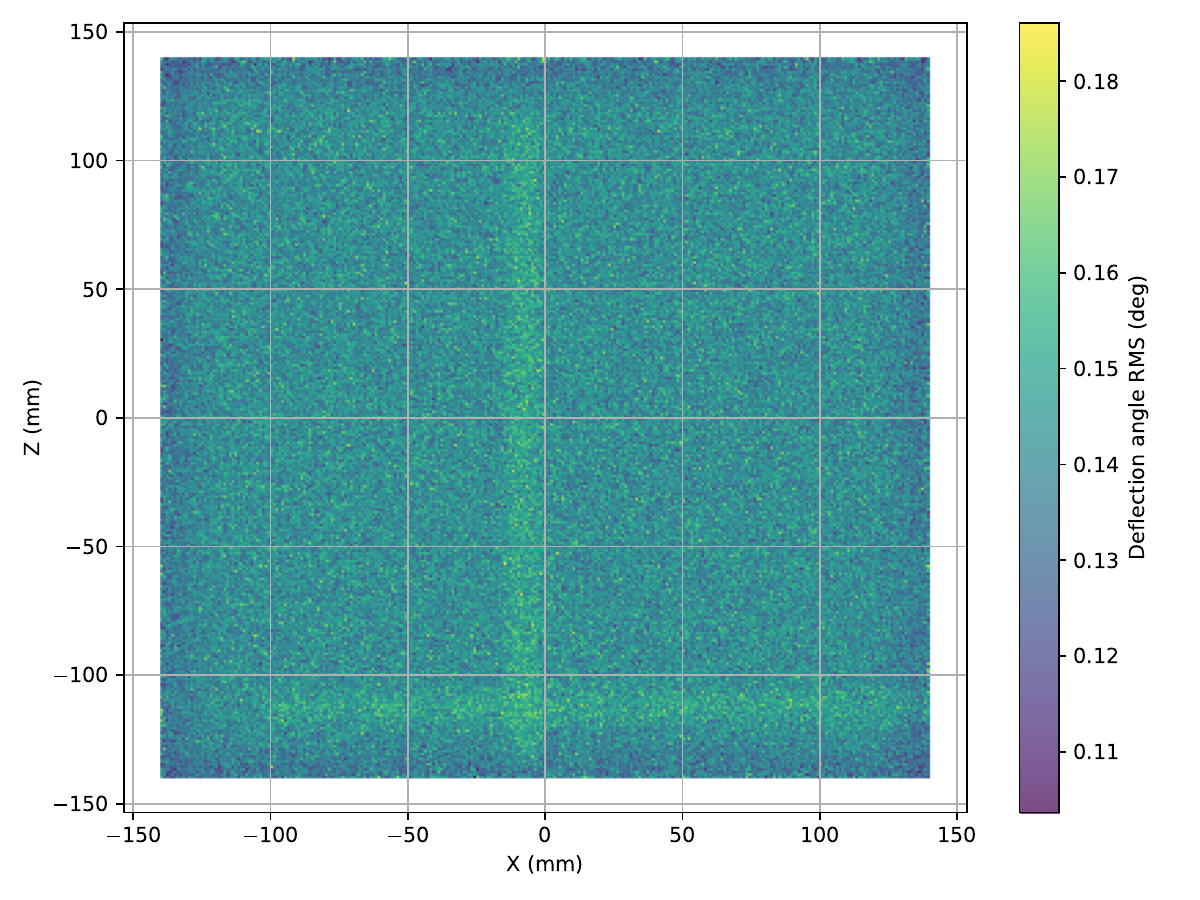}}
\fbox{\includegraphics[width=0.5\textwidth]{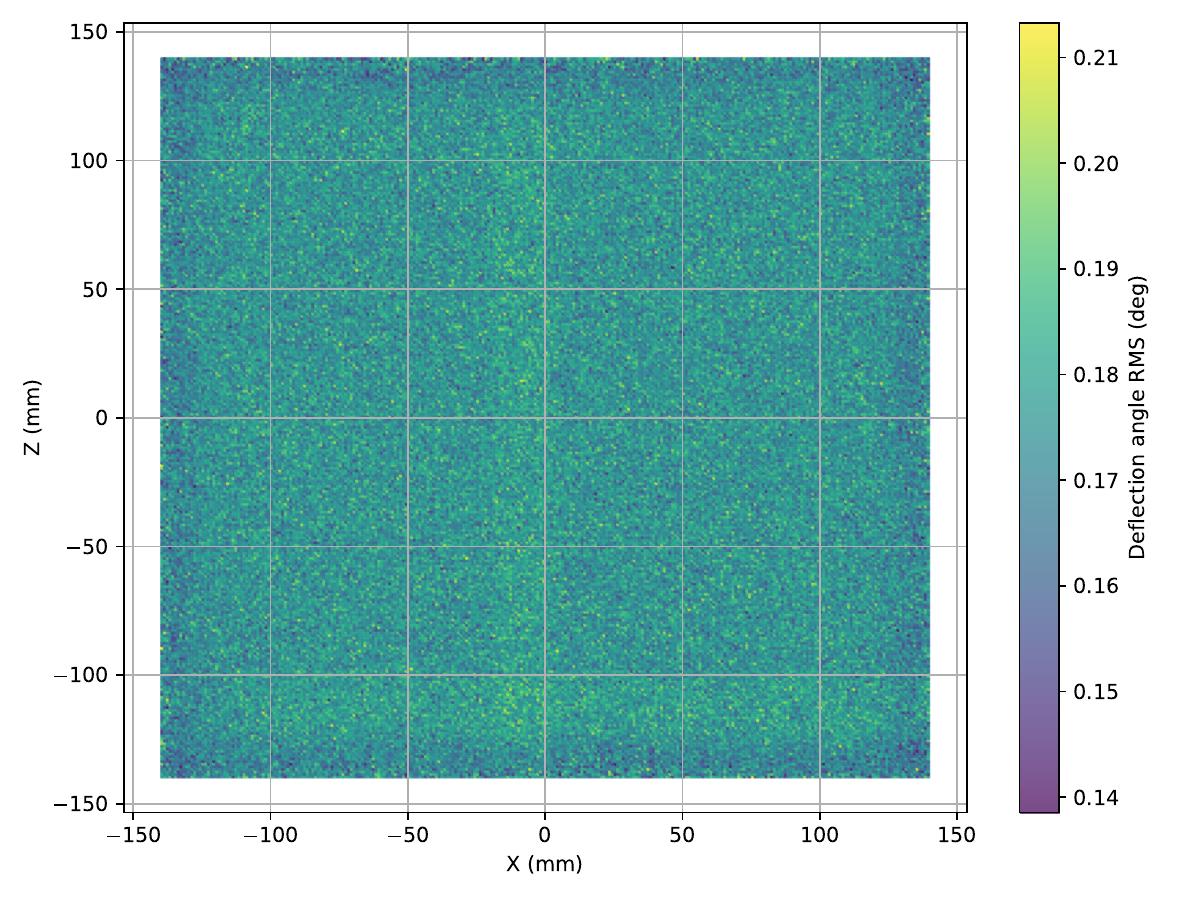}}
\fbox{\includegraphics[width=0.5\textwidth]{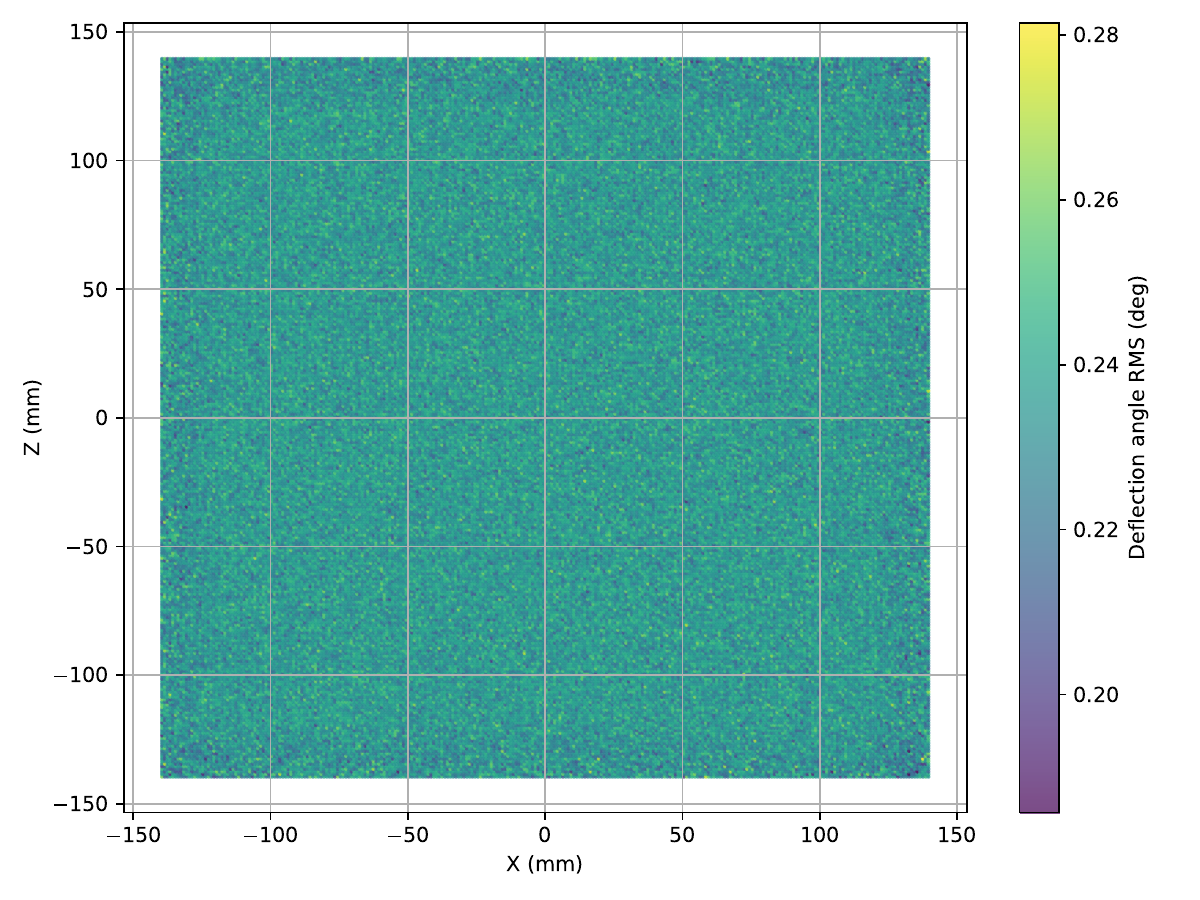}}
\caption{Scatter plots of the deflecting angle RMS of muon tracks passing through the sample geometry shown in Fig.~\ref{Geo1_1}, and travelling along the $y$-direction. The images have been created with a spatial resolution for both detecting planes of $5$ (up), $10$ (middle), and $20$ (bottom) mm; see the text for details.}
\label{Geo1_1_resolutions}
\end{figure}

Fig.~\ref{Geo1_1_resolutions} shows the imaging along the $y$-direction for the sample analysed above. We would like to stress that the inclusion of a $5$ mm smearing to planar detectors, placed at a $1$ m distance from sample front and back faces, did not affect the high-resolution capability of this technique (top). Signs of image deterioration emerged when considering a resolution of $10$ mm (middle), and no internal structures were resolvable at all with a spatial resolution of $20$ mm (bottom). It should be remarked that the spatial resolution requirement is well within the reach of several particle trackers, like large-area silicon sensors, micro-pattern gas detectors, and optical fibers-based tracking systems.

\section{Neural network algorithm development}
\label{sec:neural}

\subsection{Dataset augmentation pipeline}
\label{sec:augmentation}
The generation of synthetic data by \textsc{Geant4} is computationally intensive;
therefore, we developed a pipeline that magnifies the dataset size, by exploiting data augmentation techniques such as rotations
and reflections. Let us recall that data augmentation not only enlarges the dataset, but
also often improves the generality of the model, since it provides diverse
representations of the data. Our augmentation strategy employs specific transformations, such as rotations and reflections described below, because they preserve the underlying physics of muon interactions since muon scattering is determined by material density and composition rather than absolute orientation. All the statistical moments calculated from the muon tracking positions described below remain invariant under these transformations, ensuring that no physically inconsistent artifacts are introduced.

\begin{figure}[!ht]
\centering
\fbox{\includegraphics[width=0.8\textwidth]{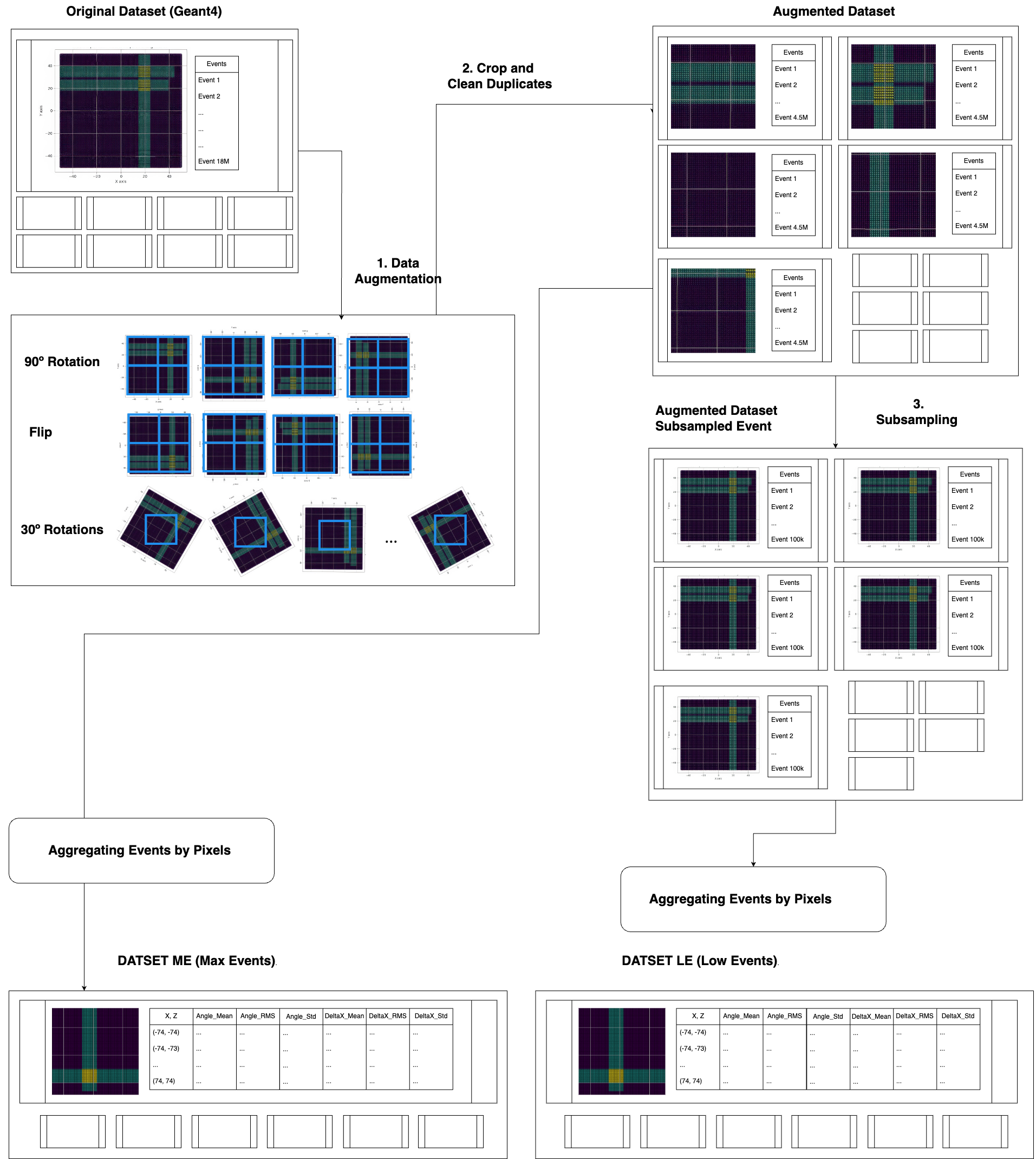}}
\caption{Data augmentation pipeline: the process begins with the original dataset $\mathscr{D}_{0}$, which undergoes data augmentation through rotations and flips to increase diversity. Then, some cropping and duplicate cleaning follow, yielding to the creation of augmented datasets with a maximum number of muon events. Subsequently, a subsampling generates datasets with fewer events, which are used to train neural networks for image reconstruction. As a final step, we proceeded to aggregate the events calculating the statistical moments of the distribution for a resolution of $140\times140$ pixels. The outcome of this pipeline consists of two dataset types: $\mathscr{D}_\textsc{me}$, in which every sample has the statistical moments arising from a distribution of $4.5\times10^6$ events, and three distinct low-event datasets ($\mathscr{D}_\textsc{le1}$, in $\mathscr{D}_\textsc{le2}$, and $\mathscr{D}_\textsc{le3}$), each containing samples whose statistical moments are derived from the aggregation of different random subsets of $10^5$ event.}
\label{Data_augmentation1}
\end{figure}

In this context, we split each element
of the dataset $\mathscr{D}$ generated by \textsc{Geant4} into four distinct
quadrants, based on the coordinates $x_{1}$ and $z_{1}$. After removing
incomplete events (i.e., those in which the muons pass through one or more detectors without activating all of them), the original dataset $\mathscr{D}_{0}$
is defined as
\begin{align}
\mathscr{D}_{0} & =\left\{ \left(\mathscr{G}_{k},\mathscr{E}_{k}\right)_{k=1,...,400}\right\} .
\end{align}
Since the dataset $\mathscr{D}$ was generated with 100 geometries,
the original dataset $\mathscr{D}_{0}$ comprises 400 geometries $\mathscr{G}_{k}$,
and each of them is associated to a collection of muon events $\mathscr{E}_{k}$,
with $k=1,...,400$. Each $\mathscr{E}_{k}$ consists of $4.5\times10^6$
muonic events, with each event recorded as four triplets $\left(x,y,z\right)$
representing the coordinates of detectors 1 through 4 crossed by the muon. Specifically, it holds that
\begin{equation}
\mathscr{E}_{k}=\left\{ \left(x_{i},y_{i},z_{i}\right)_{i=1,...,4}:\;x_{i}\in\left[-75,75\right],\:y_{i}\in\left\{ y_{1},...,y_{4}\right\},\:z_{i}\in\left[-75,75\right] \right\} .
\end{equation}
Therefore, each element of the original dataset undergoes the following transformations, as illustrated in Fig.~\ref{Data_augmentation1}:
\begin{enumerate}[label=\roman*)]
\item $90^\circ$, $180^\circ$, and $270^\circ$ rotations of the data applied to each quadrant;
\item 8 additional rotations in increments of $30^\circ$, specifically, $30^\circ$, $60^\circ$, $120^\circ$, $150^\circ$, $210^\circ$, $240^\circ$, $300^\circ$, and $330^\circ$, applied only to the central region;
\item a reflection across the vertical axis applied to each quadrant;
\item a reflection across the horizontal axis applied to each quadrant;
\item finally, a reflection through both the horizontal and vertical axes applied to each quadrant.
\end{enumerate}
This approach yields a total of 36 geometric configurations for each original geometry:
\begin{itemize}
\item 4 original quadrants
\item 12 quadrants from regular rotations (3 rotations × 4 quadrants)
\item 8 configurations from fine rotations (applied to the central region only)
\item 12 quadrants from reflections (3 reflection types × 4 quadrants)
\end{itemize}
With 100 initial geometries divided into 4 quadrants each, and 36 configurations per geometry, the Augmented Dataset $\mathscr{D}_\textsc{a}$
contains 3600 samples, each consisting of a Geometry $\mathscr{G}_{k}$ with associated events $\mathscr{E}_{k}$.
Each of these transformations was applied both to the ground truth image containing the material budget information (like the one depicted in Fig.~\ref{Geo1_1_Projections} (right)) which serves as the visual reference of the sample under test, and to the muon tracking coordinates ($x$,$y$,$z$) through the four detecting planes.

After these transformation are applied, using the ($x$,$y$,$z$) coordinates of muon tracks and assuming a given pixel size, all events corresponding to the same pixel were grouped and three distributions were created for each pixel: the displacement along $x$ (DeltaX), the displacement along $z$ (DeltaZ) and the deflection angle. From such distributions, we computed the following statistics: the mean, the RMS, the standard deviation (STD), the skewness and the kurtosis. It follows that, for each element of the augmented dataset, $15$ images (5 statistics $\times$ 3 histograms) showing such aggregated statistics were obtained.

Specifically for neural network training, two fundamental types of datasets were created each with $140\times140$ pixels: the first is $\mathscr{D}_\textsc{me}$, in which every sample has the statistical moments arising from the maximum number (i.e., $4.5\times10^6$) of events, whereas the second type consists of three low-event datasets ($\mathscr{D}_\textsc{le1}$, $\mathscr{D}_\textsc{le2}$, and $\mathscr{D}_\textsc{le3}$), each containing samples whose statistical moments are derived from a smaller number (i.e., $10^5$) of events; this entails a reduction of a factor 45 in the volume of the muonic events. Each of the three $\mathscr{D}_\textsc{le}$ datasets contains a different random subset of $10^5$ events, selected using different random seeds, to ensure robust training across different statistical realizations.
This triples the effective size of our low-event training data, resulting in 10,800 distinct training samples for the low-event regime.

It is here worth remarking that, in order to avoid any possible contamination between the training data and the test data, we purposed all the samples obtained from the geometries 1 to 50 of the dataset $\mathscr{D}$ generated by \textsc{Geant4} only for test, and moreover we confined the training only onto the samples obtained from the geometries 51 to 100 of the dataset $\mathscr{D}$.

\subsection{Neural network architecture}\label{subsec:NN_architecture}
Our neural networks employed a refined U-Net design augmented with residual-in-residual dense blocks (RRDB). The reason for this choice is that RRDB have proven to be effective in various super-resolution and image restoration tasks \cite{wang2018esrgan}. Indeed, the chosen architecture makes use of hierarchical feature extraction, and the enhanced feature gets reused in order to produce high-quality reconstructions from the complex input data generated by our \textsc{Geant4} simulations.

\begin{figure}[!ht]
\centering
\fbox{\includegraphics[width=0.2\textwidth]{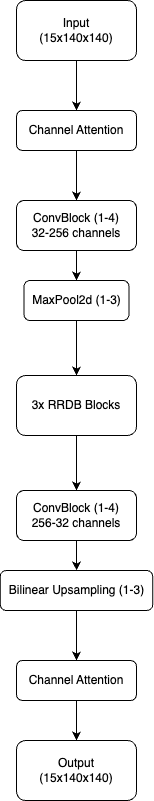}}
\caption{The architecture of our models is a U-Net architecture augmented by RRDB.}
\label{Architecture}
\end{figure}

\subsubsection{U-Net architecture with RRDB}
While muon tracking data is inherently sparse at the event level, our preprocessing approach described in Section \ref{sec:augmentation} transforms this sparse representation into dense feature maps through statistical aggregation. By calculating metrics such as mean deflection angle for each pixel position, we create dense feature maps where every pixel contains meaningful statistical information. This aggregation process addresses potential sparsity concerns and makes standard convolutional operations appropriate and efficient for our task.

The enhanced U-Net architecture consists of an encoder-decoder structure with skip connections, augmented by RRDB in the bottleneck to improve feature extraction and noise suppression. We started setting the U-Net as the baseline architecture, given its ability to handle image-to-image translation tasks through its encoder-decoder symmetry \cite{ronneberger2015u}. We should remark that the standard U-Net design is particularly suitable for capturing both global structural information (through encoding at multiple scales) and fine-grained details (via skip connections that reintroduce high-resolution features at later stages). However, we appreciated that standard U-Nets alone may provide unsatisfactory performance when applied to muographic data, which feature subtle and noisy patterns due to low muon statistics \cite{wang2018esrgan}. For this reason, we inserted RRDBs, each of which embeds multiple layers of residual dense connections, into the U-Net structure. This allowed the residual dense connections to reuse features and stabilize training while, at the same time, the residual-in-residual structure preserved spatial details with a usually better gradient flow, even in deep networks. Remarkably, this approach gave rise to an improved representational power as well as to a very good capability in dealing with scenarios characterized by a low signal-to-noise ratio. The overall architecture is depicted in Fig.~\ref{Architecture}.

\subsubsection{Detailed network configuration}
\begin{figure}[!ht]
\centering
\fbox{\includegraphics[width=\textwidth]{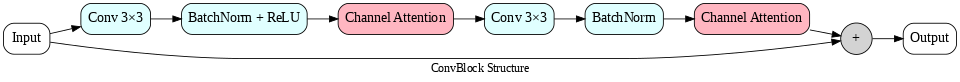}}
\caption{Convolutional block structure used within the U-Net model. The sequence of operations, including convolution, batch normalization, ReLU activation, and channel attention, is depicted.}
\label{Conv_block}
\end{figure}

The modified U-Net architecture with RRDB follows a multi-scale encoder-decoder pathway. Each encoder stage consists of a pair of convolutional layers (with batch normalization and non-linear activation), followed by a spatial downsampling operation (e.g., max pooling). Then, the decoder symmetrically upsample the compressed features back to their original resolution, thereby reintroducing high-level semantic information and merging it with spatially resolved features passed through skip connections from the corresponding encoder stage. The bottleneck of the U-Net was replaced by a sequence of RRDB. Each RRDB contains multiple dense blocks connected in a residual-in-residual fashion. These dense blocks stack several convolutional layers, Fig.~\ref{Conv_block}, and each of these layers takes as input all preceding feature maps within the block. This ultimately helps the network to learn increasingly complex representations, while reducing vanishing gradients.

\vspace{1cm}

\begin{figure}[!ht]
\centering
\fbox{\includegraphics[width=1\textwidth]{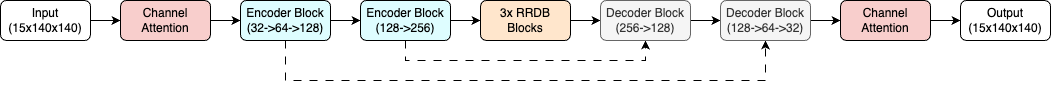}}
\caption{Detailed diagram of the encoder-decoder structure. The encoder path reduces spatial dimensions while extracting high-level features, whereas the decoder restores the spatial resolution, by integrating low-level details. }
\label{Encoder_decoder}
\end{figure}

\paragraph{Encoder and decoder paths.} The encoder starts from a set of input feature maps derived from muographic observables. It progressively reduces spatial dimensions, capturing global contextual information, see Fig.~\ref{Encoder_decoder}. Each level consists of convolutional blocks that extract local features, followed by downsampling. Conversely, the decoder receives the transformed, high-level features at the bottleneck and then refines them through successive upsampling. Each decoder stage concatenates the upsampled features with the corresponding encoder features via skip connections, thus reintroducing rich spatial details (which were lost during the downsampling).

\begin{figure}[!ht]
\centering
\fbox{\includegraphics[width=\textwidth]{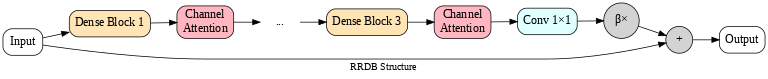}}
\caption{RRDB structure used in the bottleneck of the enhanced U-Net model. The RRDB integrates multiple dense blocks with residual connections.}
\label{RRDB}
\end{figure}

\paragraph{Connections and channel attention mechanisms.}
We have also successfully improved the noise robustness of the network, by making use of channel attention modules \cite{hu2018squeeze}. The channel attention weights feature channels based on their relative importance, and they focus the network on the critical aspects of the input data. The embedding of channel attention both within the encoding and decoding paths, as well as within the RRDB structured as in Fig.~\ref{RRDB}, causes the network to emphasize meaningful muon scattering patterns, as well as structural variations in the target object.

\subsubsection{Hyperparameter selection}
Next, we performed a systematic and detailed hyperparameter investigation. Key hyperparameters included the initial learning rate, the number of RRDB, the growth rate within the dense layers, and the choice of the normalization and activation functions. We used AdamW optimizer \cite{loshchilov2019decoupled} for a stable convergence, and we also used a learning rate scheduler with early stopping criteria, for computational efficiency and to prevent overfitting.

Table~\ref{tab:hyperparams} summarizes the key hyperparameters used in our model architecture and training process. These values were selected after extensive experimentation to optimize performance on our muographic data.

\begin{table}[h]
\centering
\caption{Summary of key hyperparameters used in our U-Net with RRDB model}
\label{tab:hyperparams}
\begin{tabular}{ll}
\hline
\textbf{Hyperparameter} & \textbf{Value} \\
\hline
Initial learning rate & $10^{-6}$ \\
Maximum learning rate & $5 \times 10^{-4}$ \\
Batch size & 16 \\
Maximum epochs & 300 \\
Early stopping patience & 30 epochs \\
Number of RRDB blocks & 3 \\
Growth rate in dense blocks & 32 \\
Optimizer & AdamW ($\beta_1 = 0.9$, $\beta_2 = 0.999$) \\
Weight decay & $10^{-5}$ \\
Gradient clipping threshold & 1.0 \\
\hline
\end{tabular}
\end{table}

\begin{figure}[!ht]
\centering
\fbox{\includegraphics[width=0.95\textwidth]{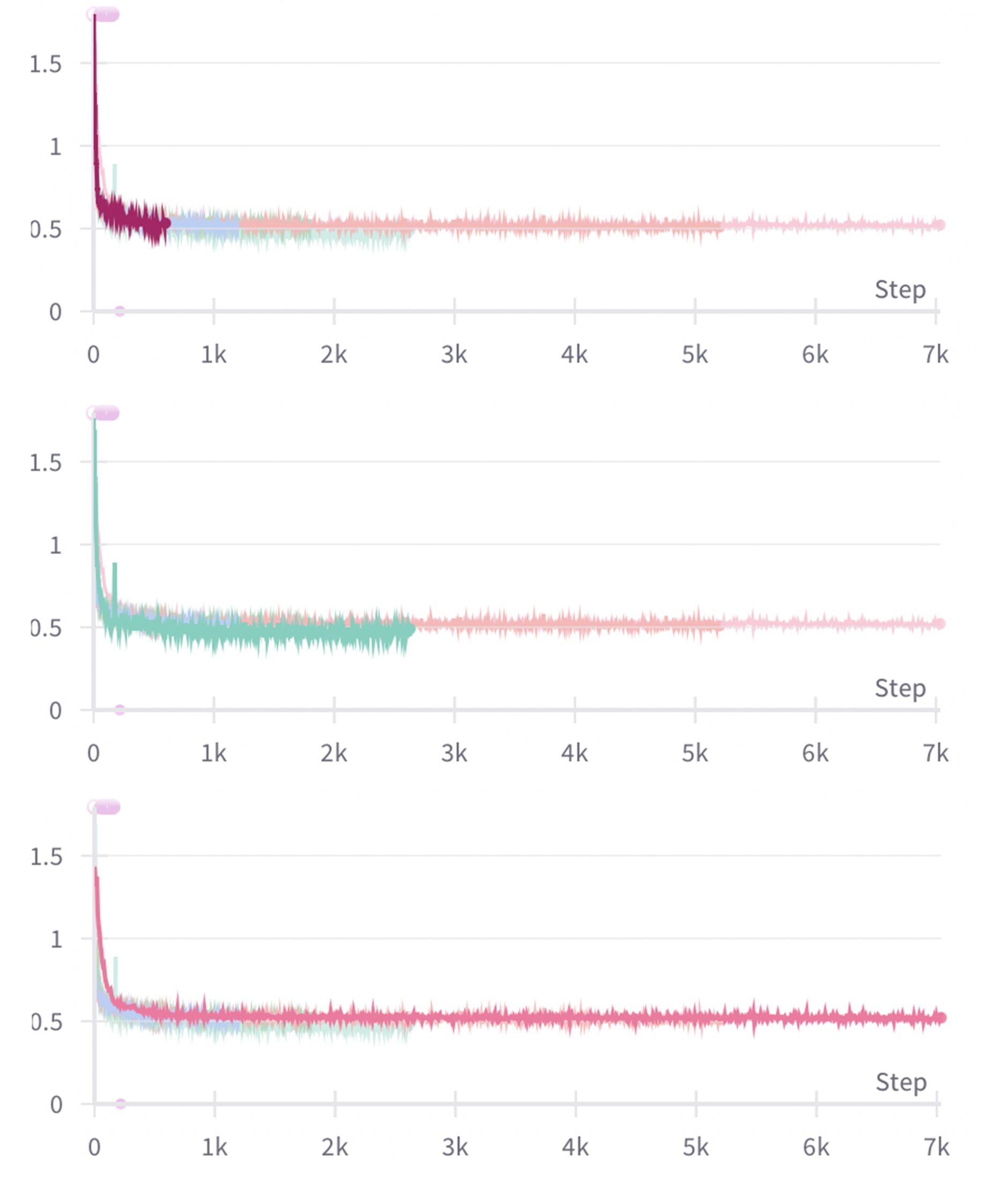}}
\caption{Batch $L_1$ loss function against training step number and its comparison across different U-Net models trained on 5, 20 and 50 geometries, respectively. One can easily appreciate a rapid decrease of the loss during the initial steps, then stabilizing around 0.5. When compared to others, the model trained with 50 geometries has a slightly lower and more stable loss.}
\label{fig:loss}
\end{figure}

\subsection{Training process}\label{subsec:training_process}
Within our methodological approach, the neural network was always trained to predict the statistics of the dataset with the largest number of events, i.e.\ $\mathscr{D}_\textsc{me}$, from the statistical moments of the dataset with the smallest number of events, i.e.\ $\mathscr{D}_\textsc{le}$. The training (conducted on a NVIDIA T4 GPU with 16 GB) was then performed onto a different number of geometries, i.e., 5, 20 and 50, with the aim to clearly showing any improvement.

\subsubsection{Loss function and optimization algorithm}\label{subsubsec:loss_opt}
For the training process we used a $L_1$ loss function, i.e., the mean absolute error (MAE) between the predicted and target images. The choice of the $L_1$ loss function can be motivated by its robust convergence and its effectiveness in determining sharp reconstructions. After the statistical aggregation process described in Section \ref{sec:augmentation}, our data is no longer sparse in the traditional sense and each pixel contains meaningful statistical moments derived from muon deflection patterns. Even "empty" regions (with no iron bars) still contain concrete material that produces measurable deflection statistics, for this reason we choose $L_1$ loss function over the mean squared error (MSE).

We also used the AdamW optimizer \cite{loshchilov2019decoupled}, which features stable convergence properties. During the entire training process, the initial learning rate (LR) was set to a low value (e.g., $10^{-6}$), and then gradually increased to its maximum (e.g., $5\times10^{-4}$). This learning rate warm-up strategy helped establish stable gradient directions during early training phases. While conventional practice often involves a warm-up followed by a decay phase, our experiments showed that the model frequently triggered early stopping before entering a substantial decay phase. This was intentional in our approach, as allowing extended training with decay often led to overfitting on our relatively limited dataset. Therefore, the effective training schedule consisted primarily of the warm-up phase, with minimal or no decay before convergence, which proved to be particularly effective for our specific data characteristics. In addition, we also applied gradient clipping, to prevent exploding gradients and to ensure stable updates during the backpropagation.

\paragraph{Training schedule and validation strategy.}
As for the training schedule, we set 300 as the maximum number of epochs, but we employed an early stopping criterion in which the validation loss does not improve for a certain number of epochs. Additionally, when the validation loss reached a plateau, we reduced the learning rate to achieve better convergence. We always maintained a 80/20 train/validation split, and the validation set was never used for weight updates, but rather it served as a checkpoint to monitor generalization and to guide hyperparametric adjustments.

\paragraph{Computational resources and performance.}
Training times varied depending on the number of geometries considered: for 5 geometries, training was approximately 1 hour, while 20 geometries took around 4 hours, and 50 geometries requested about 8 hours of runtime. The GPU memory usage remained at about 30\%.  A graphical representation of the loss during training is given in Fig.~\ref{fig:loss}.

The final trained model has a memory footprint of about 36 MB, and it features very effective inference capabilities, taking on the order of tens of milliseconds per sample.

\section{Results}\label{sec:results}
As detailed above, we considered the training on 5, 20 and 50 geometries, leaving all other geometries for testing the models. Compared to the baseline (obtained with the analytical grouping of muon events into pixels, without any machine learning technique), in all three such cases we observed a dramatic gain in the image quality. In fact, we obtained results far beyond our expectations in achieving clearer, higher-resolution images from a significantly lower number of muonic events. In order to quantify the impact of our setup, we used standard image quality metrics, such as:
\begin{itemize}
\item \textit{Peak Signal-to-Noise Ratio} (PSNR): higher PSNR indicates cleaner reconstructions and less noise.
\item \textit{Structural Similarity Index Measure} (SSIM): quantifies structural resemblance between reconstructed and target images, with values closer to 1 indicating more faithful reconstructions \cite{wang2004image}.
\item \textit{Gradient Magnitude Similarity Deviation} (GMSD): this evaluates the sharpness of the reconstructed images \cite{xue2014gradient}. Lower values indicate that edges and gradients are preserved more faithfully.
\end{itemize}

\begin{figure}[!ht]
\centering
\fbox{\includegraphics[width=1\textwidth]{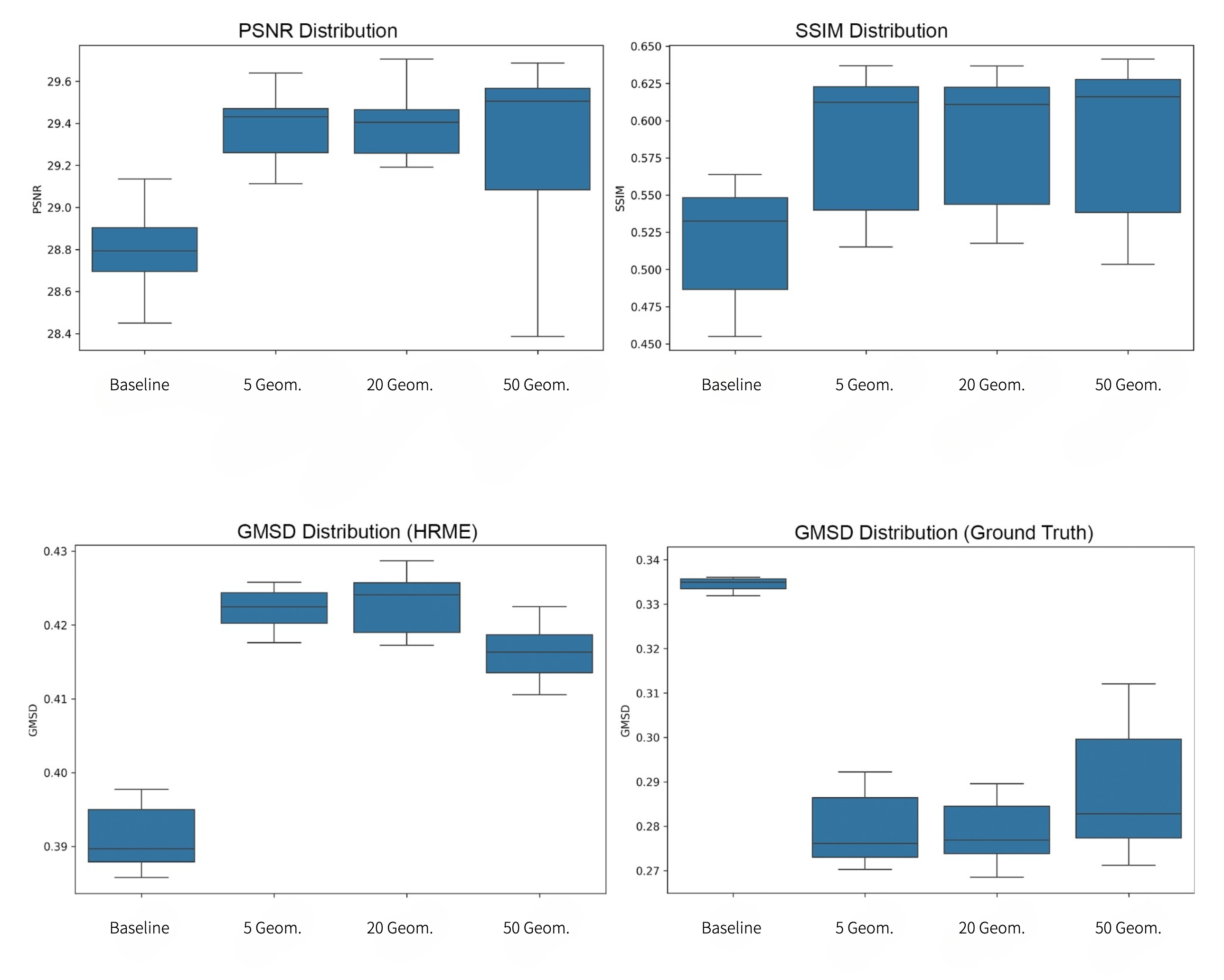}}
\caption{Distribution of the performance metrics across different neural network training setups. The top left panel shows the PSNR distribution, whose higher values indicate an improved image quality and a reduced noise. The top right panel reports the SSIM distribution, reflecting how well the predicted images preserve the structural details compared to the ground truth. The bottom left panel depicts the distribution of the GMSD pertaining to the $\mathscr{D}_\textsc{me}$ dataset, for which lower values indicate better edge preservation. Finally, the bottom right panel shows the distribution of the GMSD measured against the Ground Truth; it can be easily appreciated that the models trained on 20 and 50 geometries achieve closer similarity to the ideal image, when compared to the baseline as well as to models trained on 5 geometries.}
\label{fig:metrics}
\end{figure}

\begin{figure}[!ht]
\centering
\fbox{\includegraphics[width=0.725\textwidth]{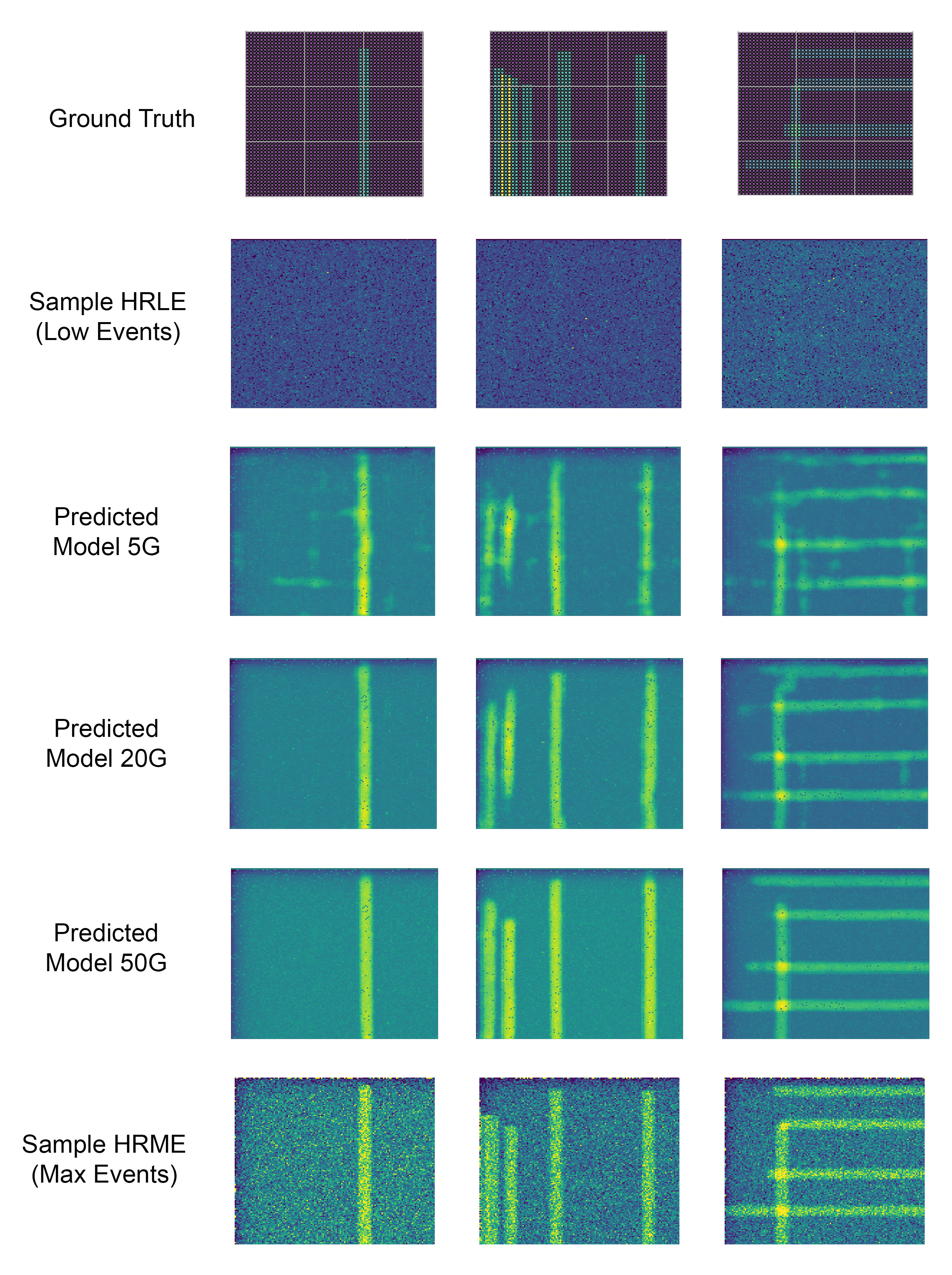}}
\caption{Ground truth, $\mathscr{D}_\textsc{le}$ sample (Low Events), predicted samples and $\mathscr{D}_\textsc{me}$ sample (Max Events) sample of Geometries 20, 21 and 27. In the first row is represented the Ground Truth with an image obtained directly from the material budget of the geometry. The second row is given by the respective samples from the  $\mathscr{D}_\textsc{le}$ dataset (Low Events, i.e.\ obtained with $10^5$ events). In the following three rows, we see the predicted images by the models trained on 5, 20, and 50 geometries, respectively. Finally, in the last row one finds the corresponding samples from the  $\mathscr{D}_\textsc{me}$ dataset (obtained with $4.5\times10^6$ events). The aggregated statistic used is the mean deflection angle.}
\label{fig:comparison}
\end{figure}

\subsection{Analysis of the quality metrics}
Considering that neural network training aims to predict statistics for a larger number of muonic events (specifically, predict the statistics of the $\mathscr{D}_\textsc{me}$ dataset, corresponding to $4.5\times10^6$ events, starting from the statistics of the $\mathscr{D}_\textsc{le}$ dataset obtained with $10^5$ events), we measured the above quality metrics on 10 geometries not included among those used for training. For each of such 10 geometries, the metrics were evaluated by comparing the results of the baseline (i.e., the statistics of $\mathscr{D}_\textsc{le}$), the predictions made with the model trained on 5 geometries (5G), on 20 geometries (20G), and on 50 geometries (50G), with the statistics obtained with the maximum number of events (i.e., the statistics of $\mathscr{D}_\textsc{me}$).

As it can be realized at a glance from Fig.~\ref{fig:metrics}, the baseline has a significantly lower PSNR and SSIM compared to the processed images with models trained on 5, 20 or 50 geometries; we recall that the  PSNR measures the quality of reconstructed images by quantifying the error between the reconstructed image and the target image (higher PSNR values correspond to lower reconstruction errors). On the other hand, the SSIM focuses on structural and perceptual similarity. On both of these metrics, the model trained on 20 geometries achieves the best compromise between the highest PSNR and SSIM values and low variability. On the other hand, the increasing of the dataset size to 50 geometries does not seem to yield to any substantial improvements on these metrics, whereas the variability seems to increase, hinting at a possible overfitting of the model.

The case of the GMSD is trickier, since this is an edge-based metric which evaluates the similarity of the gradients between images; thus, when the GMSD is lower, the edges are better preserved. Unlike the above metrics, we noticed that, as far as the GMSD is concerned, the baseline model significantly outperformed the others (Fig.~\ref{fig:metrics}, bottom left). This can ultimately traced back to the presence of a strong noise in the muonic statistics, which is considered to be a meaningful detail from the GMSD; for this reason, we decided to proceed with an additional evaluation of the images with the Ground Truth image, directly obtained by calculating the material budget (Fig.~\ref{fig:metrics}, bottom right). As expected, the GMSD of the predicted images with respect to the Ground Truth is definitely better than the one of the baseline, and this provides a non-trivial confirmation that the prediction models not only enhance the meaningful structure of the image, but they also reduce the noise. Overall, the results regarding the GMSD metric are in line with the trends observed above for the PSNR and the SSIM; this provides further evidence to the statement that the model trained with 20 geometries achieves the most consistent performance.

It is worth noting that, as will be further examined in Section \ref{sec:visualcomparison}, there exists an interesting discrepancy between the quantitative metrics discussed above and the visual quality of the reconstructed images. While the 20-geometry model demonstrates superior performance according to PSNR and SSIM metrics, visual inspection reveals that the 50-geometry model produces clearer images with better defined structures. This apparent contradiction highlights the limitations of standard image quality metrics in specialized applications like muography. The pixel-wise comparisons of PSNR and SSIM may not fully capture the perceptual quality aspects most relevant to structural detection in this context, such as edge definition and noise suppression in regions of interest. The 50-geometry model appears to better preserve the structural details critical for identifying reinforcement bars, despite showing slightly lower global metrics. This observation is partially supported by the GMSD analysis, where the 50-geometry model demonstrates an improved capability to preserve meaningful structural edges while suppressing background noise—a quality particularly valuable for accurately identifying reinforcement structures but not fully reflected in the PSNR/SSIM scores.

\subsection{Visual comparisons and qualitative analysis}
\label{sec:visualcomparison}
\begin{figure}[!ht]
\centering
\fbox{\includegraphics[width= \textwidth]{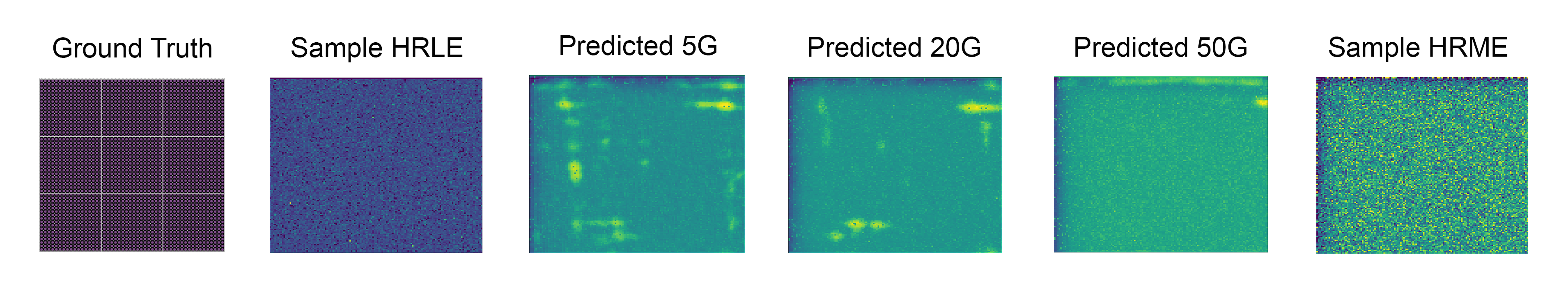}}
\caption{Ground truth, $\mathscr{D}_\textsc{le}$ sample, predicted samples and $\mathscr{D}_\textsc{me}$ sample of geometry 26, which is also a null case. As one can see, the model trained on 5 geometries introduced some artifacts, which were then reduced resp. almost completely removed by the model trained on 20 resp. 50 geometries. The aggregated statistic used is the mean deflection angle.}
\label{fig:null_case}
\end{figure}

Regardless of the statistical metrics, significant conclusions can be drawn simply by visualizing the results. In Fig.~\ref{fig:comparison}, one can see three geometries, with a representation of the material budget (Ground Truth) shown in the first row, and followed by the respective samples from the  $\mathscr{D}_\textsc{le}$ dataset (low events obtained with $10^5$ events), which are then used as input for the neural networks. In the subsequent three rows, one can see the images predicted by the models trained on 5, 20, and 50 geometries, respectively, and finally the corresponding samples from  $\mathscr{D}_\textsc{me}$ dataset (obtained with $4.5\times10^6$ events). The aggregated statistic displayed is the mean reconstructed angle.
It can be easily realized that the readability of the image greatly improves when switching from the one obtained with the statistics of $10^5$ muonic events to the one obtained with the statistics pertaining to any of the three models trained on 5, 20 or 50 geometries. Notwithstanding the aforementioned fact that the quantitative metrics tend to single out the model trained on 20 geometries as the best, the visual analysis indicates that (with a small difference) the image obtained with the model trained on 50 geometries is noticeably better. This conclusion can be confirmed by the analysis of the null case, Fig.~\ref{fig:null_case}: from the latter figure, it can be appreciated that the model trained on 5 geometries actually introduced some artifacts, which got then reduced within the model trained on 20 geometries, and further completely removed in the image obtained from the predicted statistics of the model trained on 50 geometries.

\begin{figure}[!ht]
\centering
\fbox{\includegraphics[width=\textwidth]{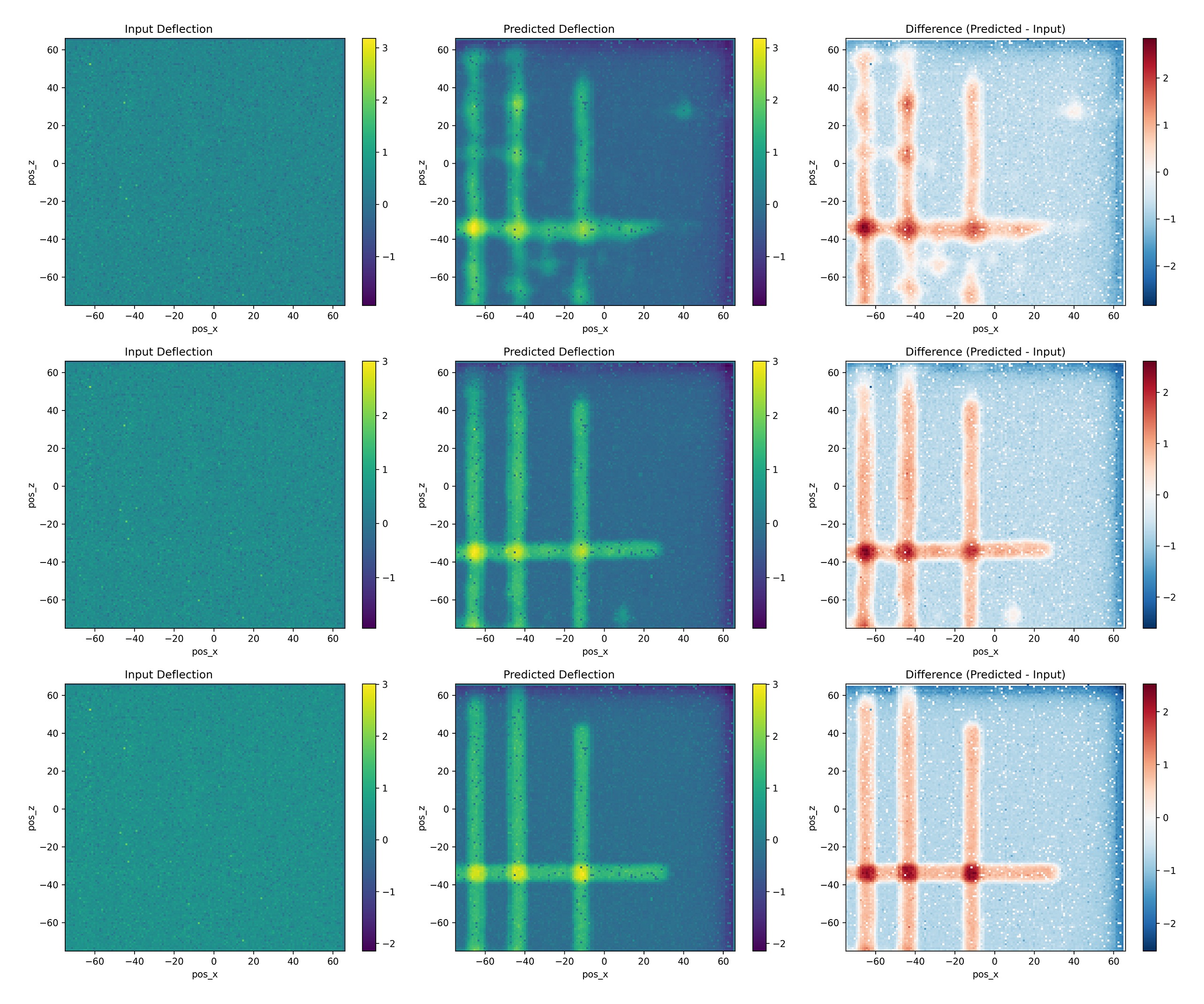}}
\caption{The first column reports the image obtained from the statistics of the sample from the Low Event dataset, i.e.\ $\mathscr{D}_\textsc{le}$, of geometry 11. On the second column, one can see the image obtained from the predicted models trained on 5, 10 and 20 geometries, from top to bottom respectively. The aggregated statistic used is the mean deflection angle. Finally, the third column reports the image resulting from the difference between the input sample and the predicted sample. }
\label{fig:difference}
\end{figure}

Fig.~\ref{fig:difference} highlights the differences between the input statistics and those resulting from the predictive models. Its first column reports the image obtained from the mean deflection angle statistic of the sample from the Low Event dataset, i.e.\ $\mathscr{D}_\textsc{le}$, whereas its second column hosts the image obtained from the predicted models trained on 5, 10 and 20 geometries, from top to bottom respectively. The colour bar quantifies the discrepancy with respect to the mean value in std units [(value-mean)/std]. Finally, in the third column one can see the image resulting from the difference between the input sample and the predicted samples. It is plain to see that the difference between the two statistics is of paramount relevance for visual understanding of the structure of the analysed geometry.

\section{Conclusion}\label{sec:conclusion}
In this work, we investigate the possible application of a new method, integrating \textsc{Geant4} simulations and advanced deep learning models, to improve the efficiency of data acquisition in muon tomography (muography) applied to reinforced concrete structures.

By performing detailed radiation-matter interaction simulations, we showed that the reconstruction of high-resolution images of typical iron support elements (of 1 cm thickness) inside 30 cm-deep concrete blocks can be achieved successfully, solely relying on analytical methods. We also showed that the requirement on the detector spatial resolution to achieve such results is well within the reach of modern tracking detectors. Additionally, through the use of convolutional neural networks, particularly U-Net architectures enhanced with residual-in-residual dense blocks (RRDB), we were able to reconstruct high-resolution images from datasets with a significantly lower number of muon events.

Deep learning models, trained on synthetic datasets generated by detailed \textsc{Geant4} simulations, have significantly improved the quality of the reconstructed images, while simultaneously reducing noise and amplifying the structural features of the images themselves. Surely, the effectiveness in the detection and visualization of internal metallic components in the concrete, even with a reduced number of muon events, represents an important step forward for the practical application of muography in structural diagnostics.

The comparative analysis of models trained on datasets of different sizes (5, 20, and 50 geometries) indicates that, although all models show improvements over the baseline, a network trained on 20 geometries suffices in demonstrating a good balance between performance and computational efficiency: this model achieves superior PSNR, SSIM, and GMSD metrics; nevertheless, we should not refrain from remarking that a visual inspection seems to single out the results of the model trained on 50 geometries as the best ones, resulting in greater image clarity, as well as in a reduction of artifact formation as the dataset size increases.

Directions for future developments can be drawn quite clearly from the present study. Just to name a few, one could use a muon distribution similar to that present in nature for cosmic muons, or also test the proposed method on some real muographic data, with the aim to validate the generalization capacity of the trained models. Furthermore, integrating a realistic detector description in our simulation suite to properly address the impact of its noise, efficiency and resolution is an ongoing activity that will be included in a new publication.

Future work should expand upon this proof-of-concept by incorporating a substantially larger dataset with more diverse reinforcement patterns, concrete compositions, and defect types. While our current approach demonstrates the feasibility of the method, a production-ready system would benefit from training on thousands of different geometries representing variations in bar thickness, spacing, orientation, concrete density, and common structural defects. Additionally, incorporating diverse muon trajectories and energy distributions that more closely match real-world cosmic ray variations would further improve generalizability for practical applications.

All in all, the present investigation lays the foundation for an effective exploitation of muography into routine civil engineering inspections, providing a powerful, non-invasive tool, capable of detecting internal structural anomalies with unprecedented and unparalleled depth and resolution. The reduction of the data acquisition times and the substantial improvement of the image quality are other two crucial features of our method, which could possibly result in more frequent and cost-effective assessments of critical infrastructures, thus contributing to improve their safety, as well as to extend their lifetime.

\bigskip

\section*{\normalsize Acknowledgments}
\vspace{-2.5pt}
We would like to thank STAP - Reabilitação Estrutural s.a.\ for its support to this research.



\begin{thebibliography}{99}

\bibitem{zhang2020muography}
Z.X.\ Zhang, T.\ Enqvist, M.\ Holma and P.\ Kuusiniemi, \textit{``Muography and its potential applications to mining and rock engineering''}, Rock Mechanics and Rock Engineering (2020) 1--15.

\bibitem{alvarez1970search}
L.W.\ Alvarez, J.A.\ Anderson, F.E.\ Bedwei, J.\ Burkhard, A.\ Fakhry, A.\ Girgis, A.\ Goneid, F.\ Hassan, D.\ Iverson, G.\ Lynch et~al., \textit{``Search for Hidden Chambers in the Pyramids. The structure of the Second Pyramid of Giza is determined by cosmic-ray absorption''}, Science \textbf{167} (1970), n.\,3919, 832--839.

\bibitem{morishima2017discovery}
K.\ Morishima, M.\ Kuno, A.\ Nishio, N.\ Kitagawa, Y.\ Manabe, M.\ Moto, F.\ Takasaki, H.\ Fujii, K.\ Satoh, H.\ Kodama et~al., \textit{``Discovery of a big void in Khufu’s Pyramid by observation of cosmic-ray muons''}, Nature \textbf{552} (2017), n.\,7685, 386--390.

\bibitem{cimmino20193d}
L.\ Cimmino, G.\ Baccani, P.\ Noli, L.\ Amato, F.\ Ambrosino, L.\ Bonechi, M.\ Bongi, V.\ Ciulli, R.\ D’Alessandro, M.\ D’Errico et~al., \textit{``3D muography for the search of hidden cavities''}, Scientific Reports \textbf{9} (2019), n.\,1, 2974.

\bibitem{tanaka2009detecting}
H.K.\ Tanaka, T.\ Uchida, M.\ Tanaka, M.\ Takeo, J.\ Oikawa, T.\ Ohminato, Y.\ Aoki, E.\ Koyama and H.\ Tsuji, \textit{``Detecting a mass change inside a volcano by cosmic-ray muon radiography (muography): First results from measurements at Asama volcano, Japan''}, Geophysical Research Letters \textbf{36} (2009), n.\,17,.

\bibitem{tanaka2013subsurface}
H.K.\ Tanaka, \textit{``Subsurface density mapping of the earth with cosmic ray muons''}, Nucl. Phys. B, Proc. Suppl. \textbf{243} (2013) 239--248.

\bibitem{tioukov2019first}
V.\ Tioukov, A.\ Alexandrov, C.\ Bozza, L.\ Consiglio, N.\ D’Ambrosio, G.\ De~Lellis, C.\ De~Sio, F.\ Giudicepietro, G.\ Macedonio, S.\ Miyamoto et~al., \textit{``First muography of Stromboli volcano''}, Scientific Reports \textbf{9} (2019), n.\,1, 6695.

\bibitem{borozdin2003radiographic}
K.N.\ Borozdin, G.E.\ Hogan, C.\ Morris, W.C.\ Priedhorsky, A.\ Saunders, L.J.\ Schultz and M.E.\ Teasdale, \textit{``Radiographic imaging with cosmic-ray muons''}, Nature \textbf{422} (2003), n.\,6929, 277--277.

\bibitem{thomay2016passive}
C.\ Thomay, J.\ Velthuis, T.\ Poffley, P.\ Baesso, D.\ Cussans and L.\ Fraz{\~a}o, \textit{``Passive 3D imaging of nuclear waste containers with muon scattering tomography''}, Journal of Instrumentation \textbf{11} (2016), n.\,03, P03008.

\bibitem{niederleithinger2022muon}
E.\ Niederleithinger, Y.\ Guangliang, D.\ Mahon and S.\ Gardner, \textit{``Muon tomography applied to assessment of concrete structures: First experiments and simulations''}, in {\em Proceedings of NDT-CE 2022}, pp.~1--8, NDT.net (2022).

\bibitem{IAEA2012}
{International Atomic Energy Agency (IAEA)}, \textit{``Muon Imaging. Present Status and Emerging Applications''}, in {\em IAEA-TECDOC-2012}, pp.~1--116, IAEA TecDoc Series (2012).

\bibitem{wang2021time}
C.\ Wang, M.\ Beer and B.M.\ Ayyub, \textit{``Time-dependent reliability of aging structures: Overview of assessment methods''}, ASCE-ASME J. Risk Uncertain. Eng. Syst. A \textbf{7} (2021), n.\,4, 03121003.

\bibitem{chen2021life}
S.\ Chen, C.\ Duffield, S.\ Miramini, B.N.K.\ Raja and L.\ Zhang, \textit{``Life-cycle modelling of concrete cracking and reinforcement corrosion in concrete bridges: A case study''}, Engineering Structures \textbf{237} (2021) 112143.

\bibitem{structural2000guideline}
{Structural Engineering Institute}, \textit{``Guideline for Structural Condition Assessment of Existing Buildings''}; American Society of Civil Engineers, Reston, Virginia (2000).

\bibitem{aci2019assessment}
{ACI Committee 562}, \textit{``Assessment, Repair, and Rehabilitation of Existing Concrete Structures - Code and Commentary''}; American Concrete Institute, Farmington Hills, USA (2019).

\bibitem{maierhofer2003nondestructive}
C.\ Maierhofer, \textit{``Nondestructive Evaluation of Concrete Infrastructure with Ground Penetrating Radar''}, Journal of Materials in Civil Engineering \textbf{15} (2003) 287--297.

\bibitem{verma2013review}
S.K.\ Verma, S.S.\ Bhadauria and S.\ Akhtar, \textit{``Review of nondestructive testing methods for condition monitoring of concrete structures''}, Journal of Construction Engineering \textbf{2013} (2013), n.\,1, 834572.

\bibitem{tanaka2023muography}
H.K.\ Tanaka, C.\ Bozza, A.\ Bross, E.\ Cantoni, O.\ Catalano, G.\ Cerretto, A.\ Giammanco, J.\ Gluyas, I.\ Gnesi, M.\ Holma et~al., \textit{``Muography''}, Nature Reviews Methods Primers \textbf{3} (2023), n.\,1, 88.

\bibitem{particle2020review}
{Particle Data Group et al.}, \textit{``Review of Particle Physics''}, Prog. Theor. Exp. Phys. \textbf{2020} (2020), n.\,8, 083C01.

\bibitem{agostinelli2003geant4}
S.\ Agostinelli, J.\ Allison, K.a.\ Amako, J.\ Apostolakis, H.\ Araujo, P.\ Arce, M.\ Asai, D.\ Axen, S.\ Banerjee, G.\ Barrand et~al., \textit{``Geant4 -- A simulation toolkit''}, Nucl. Instrum. Methods Phys. Res. A \textbf{506} (2003), n.\,3, 250--303.

\bibitem{allison2006geant4}
J.\ Allison, K.\ Amako, J.\ Apostolakis, H.\ Araujo, P.A.\ Dubois, M.\ Asai, G.\ Barrand, R.\ Capra, S.\ Chauvie, R.\ Chytracek et~al., \textit{``Geant4 developments and applications''}, IEEE Trans. Nucl. Sci. \textbf{53} (2006), n.\,1, 270--278.

\bibitem{allison2016recent}
J.\ Allison, K.\ Amako, J.\ Apostolakis, P.\ Arce, M.\ Asai, T.\ Aso, E.\ Bagli, A.\ Bagulya, S.\ Banerjee, G.\ Barrand et~al., \textit{``Recent developments in Geant4''}, Nucl. Instrum. Methods Phys. Res. A \textbf{835} (2016) 186--225.

\bibitem{wei2023esrgan}
Z.\ Wei, Y.\ Huang, Y.\ Chen, C.\ Zheng and J.\ Gao, \textit{``A-ESRGAN: Training real-world blind super-resolution with attention U-Net Discriminators''}, in {\em Pacific Rim International Conference on Artificial Intelligence}, (Singapore), pp.~16--27, Springer, 2023.

\bibitem{chen2024training}
Q.\ Chen, H.\ Li and G.\ Lu, \textit{``Training ESRGAN with multi-scale attention U-Net discriminator''}, Scientific Reports \textbf{14} (2024), n.\,1, 29036.

\bibitem{wang2018esrgan}
X.\ Wang, K.\ Yu, S.\ Wu, J.\ Gu, Y.\ Liu, C.\ Dong, Y.\ Qiao and C.\ Change~Loy, \textit{``ESRGAN: Enhanced Super-Resolution Generative Adversarial Networks''}, in {\em Computer Vision -- ECCV 2018 Workshops}, pp.~63--79, Springer International Publishing, Cham (2019).

\bibitem{ronneberger2015u}
O.\ Ronneberger, P.\ Fischer and T.\ Brox, \textit{``U-Net: Convolutional Networks for Biomedical Image Segmentation''}, in {\em Medical Image Computing and Computer-Assisted Intervention}, pp.~234--241, Springer, Cham (2015).

\bibitem{hu2018squeeze}
J.\ Hu, L.\ Shen and G.\ Sun, \textit{``Squeeze-and-Excitation Networks''}, in {\em Proceedings of the IEEE Conference on Computer Vision and Pattern Recognition}, pp.~7132--7141 (2018).

\bibitem{loshchilov2019decoupled}
I.\ Loshchilov and F.\ Hutter, \textit{``Decoupled Weight Decay Regularization''}, in {\em 7th International Conference on Learning Representations}, OpenReview.net (2019).

\bibitem{wang2004image}
Z.\ Wang, A.C.\ Bovik, H.R.\ Sheikh and E.P.\ Simoncelli, \textit{``Image quality assessment: From error visibility to structural similarity''}, IEEE Trans. Image Process. \textbf{13} (2004), n.\,4, 600--612.

\bibitem{xue2014gradient}
W.\ Xue, L.\ Zhang, X.\ Mou and A.C.\ Bovik, \textit{``Gradient magnitude similarity deviation: A highly efficient perceptual image quality index''}, IEEE Trans. Image Process. \textbf{23} (2014), n.\,2, 684--695.

\end{thebibliography}


\end{document}